\documentclass[fleqn,usenatbib]{mnras}

\usepackage[T1]{fontenc}
\usepackage{ae,aecompl}

\usepackage{graphicx}	
\usepackage{amsmath}	
\usepackage{amssymb}	
\graphicspath{{./plots/}}

\newcommand{\msun}{M$_{\odot}$}

\newcommand{\kms}{km~s$^{-1}$}
\newcommand{\ergs}{erg s$^{-1}$}

\newcommand{\HeI}{\ion{He}{I}}
\newcommand{\OI}{O~{\sc i}}
\newcommand{\Oneb}{[O~{\sc i}]}
\newcommand{\CII}{C~{\sc ii}}
\newcommand{\NaI}{Na~{\sc i}}

\newcommand{\MgI}{Mg~{\sc i}}

\newcommand{\SiII}{Si~{\sc ii}}

\newcommand{\SI}{S~{\sc i}}

\newcommand{\CaII}{Ca~{\sc ii}}
\newcommand{\TiII}{Ti~{\sc ii}}

\newcommand{\FeII}{Fe~{\sc ii}}

\newcommand{\Fefs}{$^{56}$Fe}
\newcommand{\Cofs}{$^{56}$Co}
\newcommand{\Nifs}{$^{56}$Ni}
\newcommand{\mej}{$M_\mathrm{ej}$}
\newcommand{\ek}{$E_\mathrm{k}$}
\newcommand{\vph}{$v_\mathrm{ph}$}

\newcommand{\trise}{$t_{-1/2}$}
\newcommand{\tdecay}{$t_{+1/2}$}
\newcommand{\eom}{$E_\mathrm{k}/M_\mathrm{ej}$}
\newcommand{\lam}{$\lambda$}

\newcommand{\Eh}{$E\left(B-V\right)_\mathrm{host}$}
\newcommand{\Emw}{$E\left(B-V\right)_\mathrm{MW}$}
\newcommand{\Etot}{$E\left(B-V\right)_\mathrm{tot}$}

\newcommand{\mni}{$M_\mathrm{Ni}$}

\newcommand{\tp}{$t_\mathrm{p}$}

\newcommand{\mzams}{$M_\mathrm{ZAMS}$}

\newcommand{\texp}{$t_\mathrm{exp}$}

\newcommand{\lph}{$L_\mathrm{ph}$}
\newcommand{\ScII}{Sc~{\sc ii}}

\newcommand{\mhe}{$M_\mathrm{He}$}

\title[SN2017ein]{Observations and Spectral Modelling of the Narrow--Lined Type Ic SN~2017ein}

\author[J.~J.~Teffs]{J.~J.~Teffs$^{1}$\thanks{E-mail:j.j.teffs@ljmu.ac.uk}, S.~J.~Prentice$^{2}$, P.~A.~Mazzali$^{1}$, C. Ashall$^{3}$
\\
$^{1}$Astrophysics Research Institute, Liverpool John Moores University, IC2, Liverpool Science Park, 146 Brownlow Hill, \\  Liverpool L3 5RF, UK\\
$^{2}$School of Physics, Trinity College Dublin, The University of Dublin, Dublin 2, Ireland\\
$^{3}$Institute for Astronomy, University of Hawai'i at Manoa, 2680 Woodlawn Dr.Hawai'i, HI 96822, USA
}

\date{Accepted XXX. Received YYY; in original form ZZZ}

\pubyear{2021}

\begin{document}
\label{firstpage}
\pagerange{\pageref{firstpage}--\pageref{lastpage}}
\maketitle

\begin{abstract}
SN~2017ein is a narrow--lined Type Ic SN that was found to share a location with a point--like source in the face on spiral galaxy NGC 3938 in pre--supernova images, making SN~2017ein the first credible detection of a Type Ic progenitor. Results in the literature suggest this point--like source is likely a massive progenitor of 60--80 \msun, depending on if the source is a binary, a single star, or a compact cluster. Using new photometric and spectral data collected for 200 days, including several nebular spectra, we generate a consistent model covering the photospheric and nebular phase using a Monte Carlo radiation transport code. Photospheric phase modelling finds an ejected mass 1.2--2.0 \msun\ with an \ek\ of $\sim(0.9 \pm0.2)\times 10^{51}$ erg, with approximately 1 \msun\ of material below 5000 \kms\ found from the nebular spectra. Both photospheric and nebular phase modelling suggests a \Nifs\ mass of 0.08--0.1 \msun. Modelling the [\OI] emission feature in the nebular spectra suggests the innermost ejecta is asymmetric. The modelling results favour a low mass progenitor of to 16--20 \msun\, which is in disagreement with the pre--supernova derived high mass progenitor. This contradiction is likely due to the pre--supernova source not representing the actual progenitor.

\end{abstract}




\section{Introduction}

Type Ic SNe are a sub-class of core collapse supernovae in which the progenitor star has lost all of its H envelope and all or almost all of its He envelope prior to collapse. The processes needed to strip this material is thought to be periods of significant mass loss driven by winds (e.g. \citet{Nomoto1995,Langer2012}) or through some binary interaction mechanism. For high mass single star evolution, the LBV phase may be responsible for the significant mass loss that could produce H/He-stripped SNe, but this may not be able to reproduce low mass events such as SN1994I \citep{Nomoto1994, Sauer2006}. However, single star evolutionary models are often unable to effectively strip all the H required to produce Type Ib SNe and all the H/He required to produce the Type Ic SNe. Mass loss during a common envelope phase or Roche lobe mass transfer for stars evolving in a binary can explain stripping but can leave thin shells of He of low mass above the carbon and oxygen rich (CO) core \citep{Nomoto1995, Yoon2010}. This may explain the sometimes observed weak \HeI\ lines in Type Ic SNe \citep{Elmhamdi2006,Prentice2018}.

Several Type II SNe progenitors have been likely identified, summarised in \citet{Smartt2009}, and these observations have given some constraints on the possible progenitor mass distribution for Type II SNe, but no Type Ib/c progenitors have been definitively observed yet. The Type II identifications have suggested an upper limit for the observed red supergiants (RSG) to be near $\sim$17 \msun, while the theoretical upper limit was expected to be closer to 20-25 \msun. This difference between the observed and theoretical upper limit is called the ``RSG problem" \citep{Smartt20092}. This gap may be related to a poor understanding of late time stellar evolution and mass loss rates but is still an open question \citep{Davies2018,Davies2020}. For Type Ib/c SNe, the ejecta mass distribution is more varied compared to non-stripped SNe as shown in \citet{Prentice2019} and estimating the progenitor mass of these SNe from ejecta masses also results in a wide range of possible progenitor masses. In addition, the binary and single star channels for Type Ib/c SNe may contribute to the total number of Type Ib/c SNe at different \mzams.

SN~2017ein was first detected by Ron Arbour on 2017 May 25.97 \citep{ArbourTNS}, in the face on spiral galaxy NGC 3938. The object was classified soon after detection as an early Type Ic SN \citep{2017TNSCR.599....1X}. After the detection, \citet{VanDyk2018, Kilpatrick2018, Xiang2019} used archival Hubble images and detected a point-like source at the same location of SN~2017ein, which would represent the first progenitor detection for a Type Ic SN. All three groups used a combination of methods to derive an estimated progenitor mass based on the observations of this point source. One method used is to calculate the colour, derived from observations using the two Hubble filters F555W and F814W, and compare these values to a set of single and binary star evolutionary tracks. \citet{VanDyk2018} used a set of evolved rotating massive stars from MIST \citep{mist} and a set of binary star evolution models from BPASS V2.1 \citep{bpass1} and estimated a single star \mzams\ of $\sim$47--48 \msun\ if the star evolved alone and a \mzams\ of $\sim$60--80 \msun if the star evolved in a binary, depending on the metallicity of the stars. \citet{Xiang2019} used a set of evolved rotating massive stars from \citet{georgy2012} and estimate a \mzams\ of $\sim$ 60 \msun. Assuming the point-like source was instead a compact blue cluster, all three groups estimate the \mzams\ to within the same mass range of the previous estimates using single or binary star conclusions. Using the nearby environment to constrain the properties of the star, under the assumption of a single phase of star formation, they again find a similar progenitor mass.

The semi-analytic Arnett fits \citep{Arnett1982} for \Nifs\ powered SNe used in \citet{VanDyk2018} give an estimated \mej\ of 1-2 \msun\, while a similar model in \citet{Xiang2019} finds a \mej\ of 0.9 $\pm$ 0.1 \msun. Both recognise that the estimated progenitor mass is far more massive than what the \mej\ estimates would suggest. \citet{Maund2016} estimated a \mzams\ for the progenitor of SN~2007gr, using similar methods as above, to be $\sim$40 \msun. This is in contrast to the progenitor mass inferred from the low mass ejecta modelled from the nebular spectra by \citet{Mazzali2010} of approximately $\sim$1 \msun, which would be analogous to a $\sim$15 \msun\ progenitor. 

In this work, we first present and discuss observations of SN~2017ein and the data reduction methods in Section \ref{sec:obs_meth}, the resulting photometric light curves in Section \ref{sec:lc}, and the observed spectra in Section \ref{sec:obs_spec}. Using this data, we model both the photometric and nebular phase of the spectral evolution and generate a bolometric light curve in Sections \ref{sec:syn_mod}. In Section \ref{sec:discus}, we discuss the synthetic model and its results with respect to stellar evolution and possible progenitor proprieties.



\section{Data collection}\label{sec:obs_meth}

Observations of SN~2017ein began on 2017 May 28.91 (MJD 57901.91) with IO:O on the Liverpool Telescope (LT) \citep{Steele2004}, based at the Roque de los Muchachos Observatory, La Palma, Spain.
The first LT spectroscopic observation was made with the SPectrograph for the Rapid Acquisition of Transients \citep[SPRAT;][]{Piascik2014} on 2017 May 29.04 (MJD 57902.04). 
Photometry was performed on sources in the exposures using a custom {\sc python} pipeline, which called {\sc pyraf} to run standard {\sc iraf} routines. The instrumental magnitudes were then calibrated to Sloan Digital Sky Survey \citep[SDSS;][]{Ahn2012} standard stars in the field.
SPRAT spectra were automatically reduced and the  wavelength and flux calibration was done via the LT pipeline \citep{Barnsley2012} and a custom {\sc python} script.
The spectra were then flux calibrated to the photometry.

\begin{table}
	\caption{Several relevant observational properties of SN\,2017ein}
    \begin{tabular}{ll}
    \hline
    $\alpha$ & 11:52:53.25\\
    $\delta$ (J2000) & +44:07:26.20 \\
    Host & NGC 3938 \\
    $\mu$ & 31.5 mag \\
    \Emw & 0.019 mag\\
    \Eh & 0.4 mag$^{\dag}$\\
    \hline
    $^{\dag}$ \citet{VanDyk2014,Xiang2019}
    \end{tabular}
    \label{tab:props}
\end{table}

\subsection{NGC 3938 - distance and reddening} \label{subsec:distance_redd}
The distance to the host galaxy, NGC 3938 is highly uncertain, and this has been discussed at length in both \citet{VanDyk2018} and \citet{Xiang2019}.
The latter adopts $\mu=31.38\pm{0.3}$ mag, while the former takes a range between 31.15 to 31.75 mag.
We adopt the average of the average value between recent Tully-Fisher measurements of the distance toward NGC 3938\footnote{NED}, giving $\mu=31.5$ mag.
The dust maps of \citet{Schlafly2011} indicate that extinction in the direction of NGC 3938 is \Emw $=0.019$ mag. 

Reddening at the source has been extensively discussed in the literature.
Estimating \Eh\ from the host \NaI\ D lines in our spectra gives a large range of values from 0.1 to 0.5 mag \citep{Poznanski2012}, but these spectra are low resolution, thus we defer to the value of \Eh $= 0.4$ mag given in \citet{VanDyk2018} and \citet{Xiang2019} for their higher resolution spectra.


\section{Light Curves} \label{sec:lc}

\begin{figure*}
	\centering
	\includegraphics[scale=1.3]{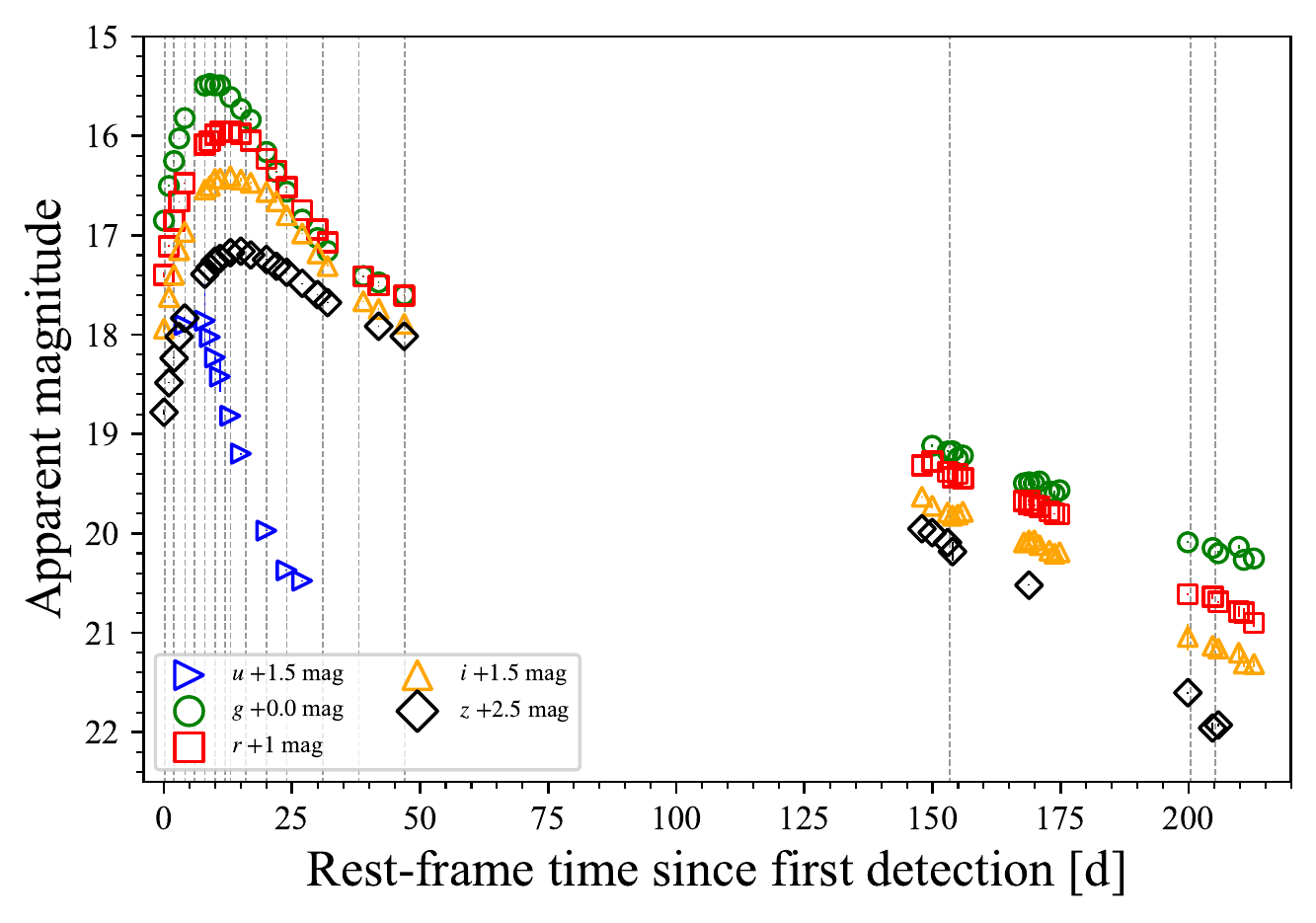}
	\caption{ The LT multi-colour light curves of SN\,2017ein. Dashed grey lines represent epochs of LT:SPRAT spectroscopy  }
	\label{fig:lc}
\end{figure*}

\begin{table*}
    \centering
    \caption{A sample of the Liverpool Telescope photometry. The entire table is available online in machine readable format via WISeREP.}
    \begin{tabular}{lccccc}
    \hline
    MJD & $u$ & $g$ & $r$ & $i$ & $z$ \\
       & [mag] & [mag] & [mag] & [mag] & [mag] \\
    \hline   
57901.91 & $-$ & $16.85\pm{0.04}$ & $16.40\pm{0.03}$ & $16.44\pm{0.04}$ & $16.28\pm{0.03}$ \\
57902.88 & $-$ & $16.50\pm{0.03}$ & $16.11\pm{0.02}$ & $16.12\pm{0.02}$ & $15.98\pm{0.04}$ \\
57903.89 & $-$ & $16.25\pm{0.03}$ & $15.85\pm{0.02}$ & $15.90\pm{0.04}$ & $15.74\pm{0.03}$ \\
57904.89 & $-$ & $16.02\pm{0.02}$ & $15.66\pm{0.02}$ & $15.65\pm{0.02}$ & $15.52\pm{0.03}$ \\
57905.97 & $16.40\pm{0.04}$ & $15.82\pm{0.03}$ & $15.47\pm{0.04}$ & $15.46\pm{0.03}$ & $15.33\pm{0.04}$ \\

    \hline
    \end{tabular}
    
    \label{tab:photometry_table}
\end{table*}

Figure~\ref{fig:lc} shows the $ugriz$ light curves, given in Table~\ref{tab:photometry_table}, obtained with the LT over $\sim$ 240 days.
SN~2017ein was discovered early, as evidenced by the rapid rise in the light curves. Only the rise of $u$-band is missed.
As is typical for SE--SNe, the light curves peak progressively later in the redder bands.
The peak of the $u$-band is 3--5 days before $g$-band, which is longer than the $\sim1$ day measured for similar objects \citep{Taddia2018}.
The transient went behind the sun just as the light curves settled on to the decay tail. 

\begin{figure}
    \centering
    \includegraphics[scale=0.6]{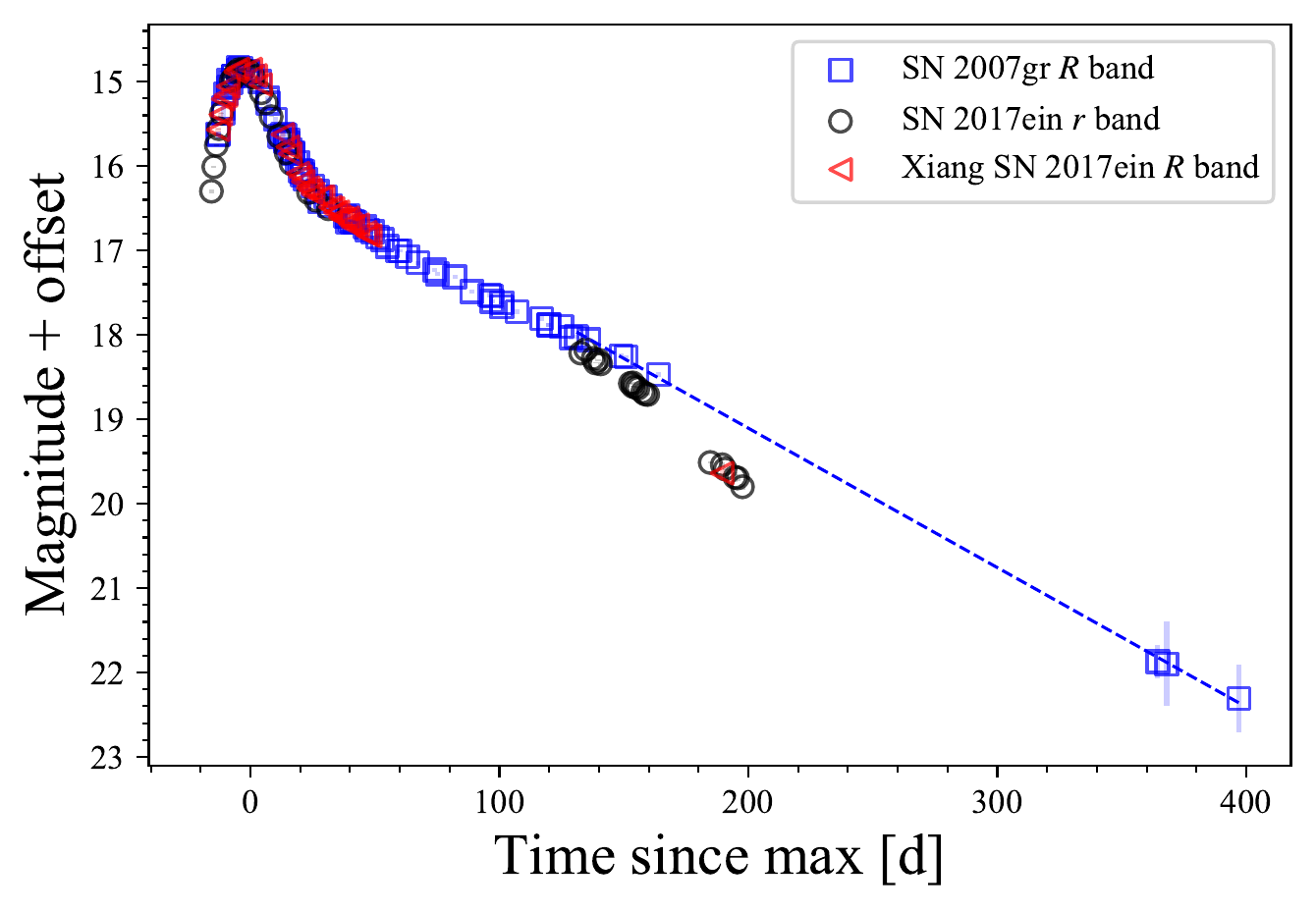}
    \caption{Comparison between our $r$-band light curve of SN\,2017ein (black circles), the \citet{Xiang2019} $R$-band light curve for the same object (red triangles) and the $R$-band light curve of the spectroscopically similar SN\,2007gr (blue squares). The late time light curves of SN\,2017ein deviate from SN\,2007gr after 100 days. This drop off is seen even more prominently in the $i$- and $z$-bands; see Fig~\ref{fig:lc}. A linear fit between the gap in observations is included to guide the eye. It shows that the later decline in SN\,2007gr is steeper than at around 100 days, but it is clearly not as pronounced as in SN\,2017ein. }
    \label{fig:rbands}
\end{figure}

A curious behaviour is seen in the later $z$-band evolution, which shows a dramatic decay in luminosity away from the linear decline normally seen in these types of events. Further investigation reveals that this is also seen to a lesser extent in $r$- and $i$-bands.
The drop-off in SN\,2017ein is pronounced when shown against the evolution of SN\,2007gr, which is demonstrated in Fig.~\ref{fig:rbands} for the $r$- and $R$-band light curves of SN\,2017ein, the latter from \citep{Xiang2019} and the $R$-band of SN\,2007gr \citep{Hunter2009}.
It can be see that for the first $\sim$ 100 days the three light curves track each other but by the time the observations of SN\,2017ein were resumed there is a clear deviation.
As the decline is seen in both SN\,2017ein data sets we can conclude that this is intrinsic to the event and is not a consequence of a data reduction issue.

\subsection{Pseudo-bolometric Light Curve} \label{sec:bol}
The $ugriz$ photometry allows the construction of a pseudo-bolometric light curve with SEDs covering 3000 -- 10000 \AA.
It is constructed using the observed photometry, corrected for \Etot\ using the extinction law of \citet{CCM} and $R_v=3.1$, which is then converted to a flux and the spectral energy distribution for each data integrated over the wavelength range. Finally, the bolometric luminosity was calculated using the luminosity distance derived from the distance modulus.
$u$-band magnitudes were estimated for epochs when no $u$-band observations were taken by assuming a constant $u-g$ band colour for dates prior to the first observation and after the final $u$-band observation.
The method is described in more detail in \citet{Prentice2016}.

\begin{figure}
    \centering
    \includegraphics[scale=0.67]{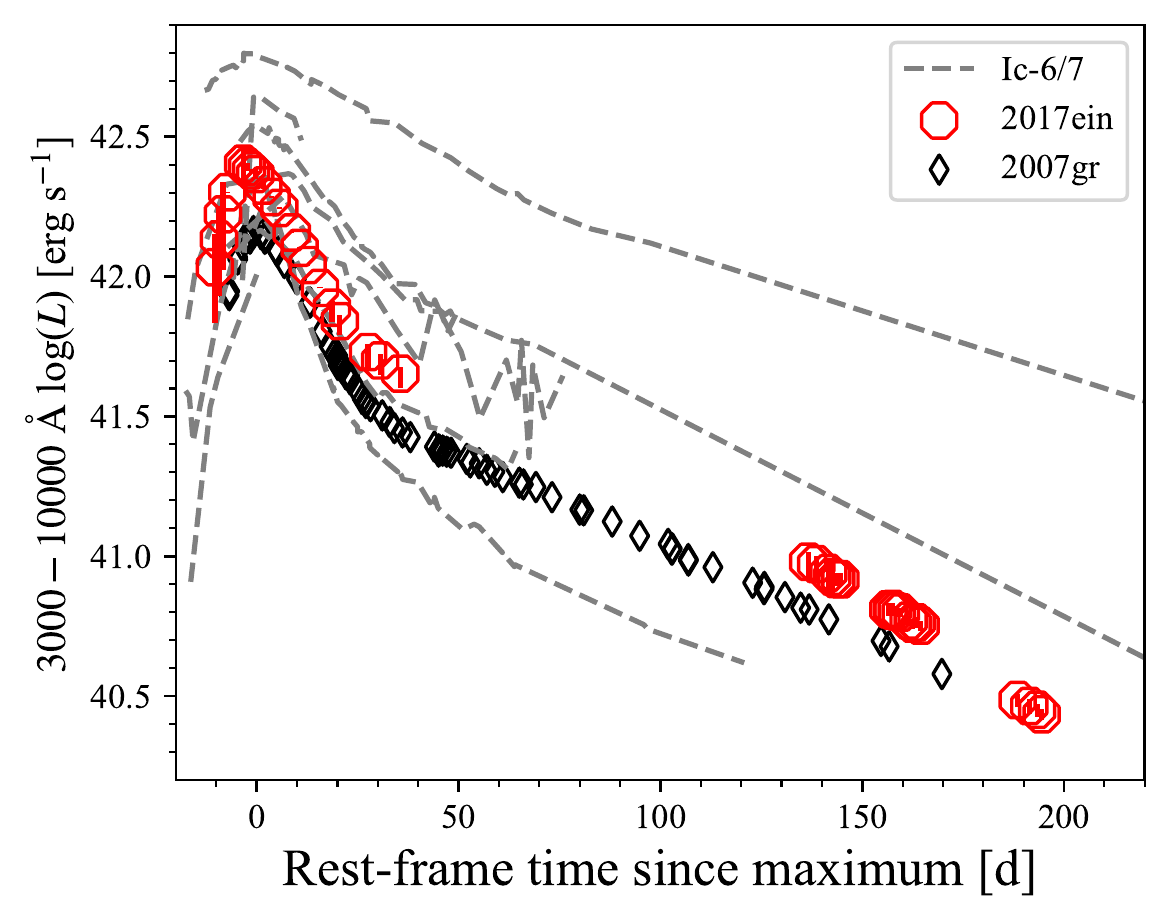}
    \caption{The pseudo-bolometric light curve of SN 2017ein (red circles) compared with SN~2007gr (black diamonds) and a sample of SNe Ic-6/7 (Grey dashed lines) \citep{Prentice2016}. }
    \label{fig:CompareBolometrics}
\end{figure}

The pseudo-bolometric light curve is shown in comparison with a sample of SNe Ic in Fig.~\ref{fig:CompareBolometrics}. 
When looking at the light curve properties in the photospheric phase, SN~2017ein is a typical SN Ic-7 \citep[a narrow lined SN Ic as per][]{Prentice2017}.
The peak luminosity for the pseudo-bolometric light curve is $(2.5\pm{0.3}) \times 10^{42}$ \ergs\ and is typical for SNe Ic-6/7.
The characteristic time-scales, \trise\ and \tdecay\ measure the rise from luminosity at half maximum to maximum, and from maximum to half maximum respectively. 
We find \trise\ $= 9.2\pm{1}$ d and \tdecay\ $=15.4\pm{0.4}$ d\footnote{These values agree with those measured from a $4000-10000$ \AA\ pseudo-bolometric light curve \citep{Prentice2016,Prentice2019} to within the uncertainties of the measurements. }.
These values compare favourably with the SN Ic-6/7 medians of $9\pm{3}$ d and $15\pm{2}$ d respectively \citep{Prentice2016,Prentice2019}.

It has been established that in most respects SN~2017ein is a standard SN Ic. However, the unusual decay in the $r$-, $i$-, and $z$-bands leads to the bolometric light curve deviating from a linear decay line. This is seen in Fig.~\ref{fig:CompareBolometrics} as the slope of SN~2017ein light curve gets increasingly steeper with time, exceeding that of the comparison objects. 

\section{Spectra}\label{sec:obs_spec}
\begin{figure}
	\centering
	\includegraphics[scale=0.6]{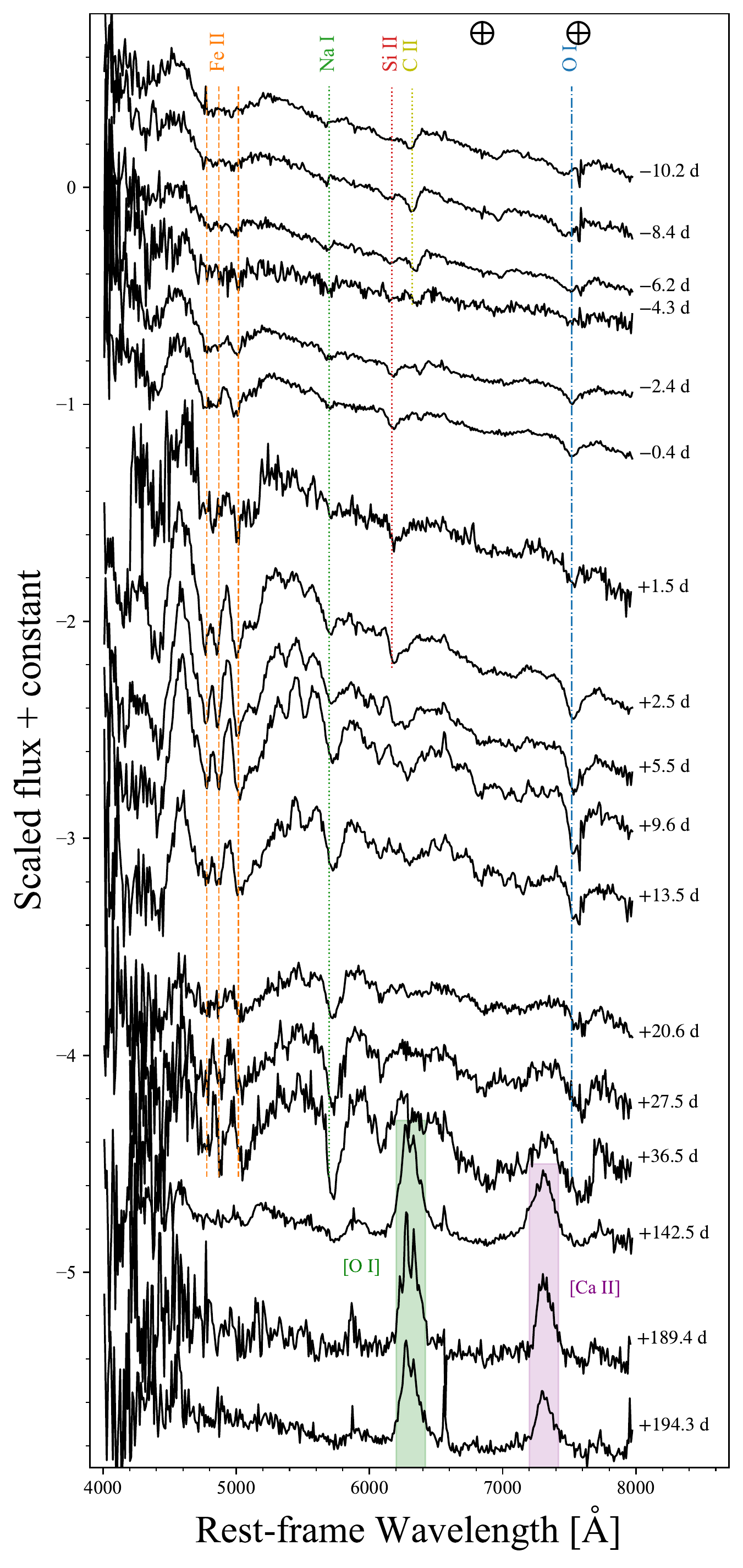}
	\caption{The flux calibrated spectra of SN 2017ein as obtained by LT:SPRAT. Epochs are since bolometric maximum.}
	\label{fig:spectra}
\end{figure}

The LT spectra are shown in Fig.~\ref{fig:spectra}.
The first few spectra show clear, but narrow absorption features. The \FeII\ dominated region around 5000 \AA\ is clearly blended.
Over time the \FeII\ \lam\lam4924, 5018, 5169 separate and become strong.
The \NaI\ D line evolves from a weak absorption at $-10.2$ d to become the strongest line in the optical spectra at around two weeks after the time of the light curve peak, \tp.
\OI\ \lam 7774 is visible blueward of the Telluric feature at $\sim7600$ \AA\ and remains so until the velocity of the line decreases and the features merge two weeks after maximum light.
\SiII\ \lam 6355 is not strong except for a few days around \tp, the first clear indication of this line is at $-8.4$ days. Shortly after \tp\ the spectra become more complex in this region, with more absorption lines appearing. 

The SN falls into the Ic-7 He-poor subgroup under the classification of \citet{Prentice2017} because the mean number of features at \trise\ and \tp\ is 7. 
This group is mainly defined by the separation of \FeII\ \lam\lam4924, 5018, 5169 into three distinct components.
This sub-class has diverse photometric properties, and includes SN~2011bm \citep{Valenti2012}, which has a rise time of $>35$ days.

\subsection{Line velocities}
\begin{figure}
	\centering
	\includegraphics[scale=0.6]{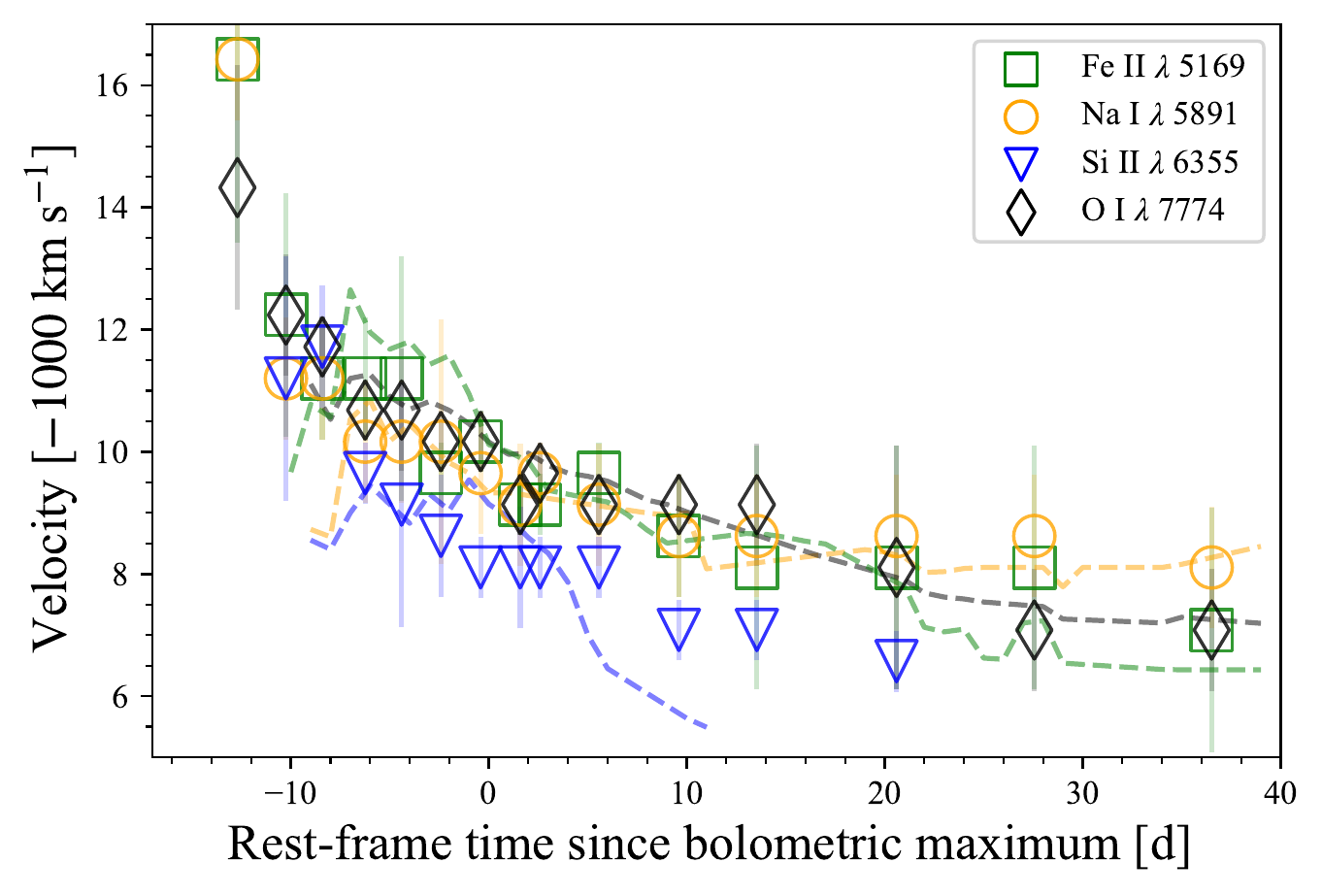}
	\caption{Velocities as measured from the absorption line minima. Also shown as dashed lines are the median line velocities for a selection of SNe Ic 6/7.  }
	\label{fig:vels}
\end{figure}

Figure~\ref{fig:vels} gives the line velocities of SN 2017ein as a function of time for \FeII\ \lam5169, \NaI\ D, \SiII\ \lam6355, and \OI\ \lam7774.
Line velocities are measured from the centre of the absorption profile minimum, the uncertainty represents the width in velocity space of this minimum.
The highest velocities are those in the classification spectrum \citep{2017TNSCR.599....1X}, at $-12.8$ d, and within a few days of explosion. It displays \FeII\ and \NaI\ at $\sim16000$ \kms\ while \OI\ is at $v\sim14000$ \kms. 
This spectrum is the only one accessible to this work where the wavelength range covers the \CaII\ NIR triplet, for which a velocity of $19000\pm{3000}$ \kms is measured.

The line velocities decrease rapidly over the course of a week, which further hints at the young nature of the transient.
Comparison with the mean line velocities for a sample of Ic-6/7 SNe \citep{Prentice2019} shows that the velocity curves are typical for a SE-SN of this type.
The velocity curve also demonstrates that the line forming regions for each element overlaps, suggesting some degree of mixing within the ejecta. 
Some stratification is most noticeable between \SiII\ \lam 6355 and the other elements.
After maximum light the velocities all level off to between $\sim 7000-9000$ \kms. 

\subsection{Similar SNe} \label{subsec:similar_sne}

\begin{figure}
	\centering
	\includegraphics[scale=0.45]{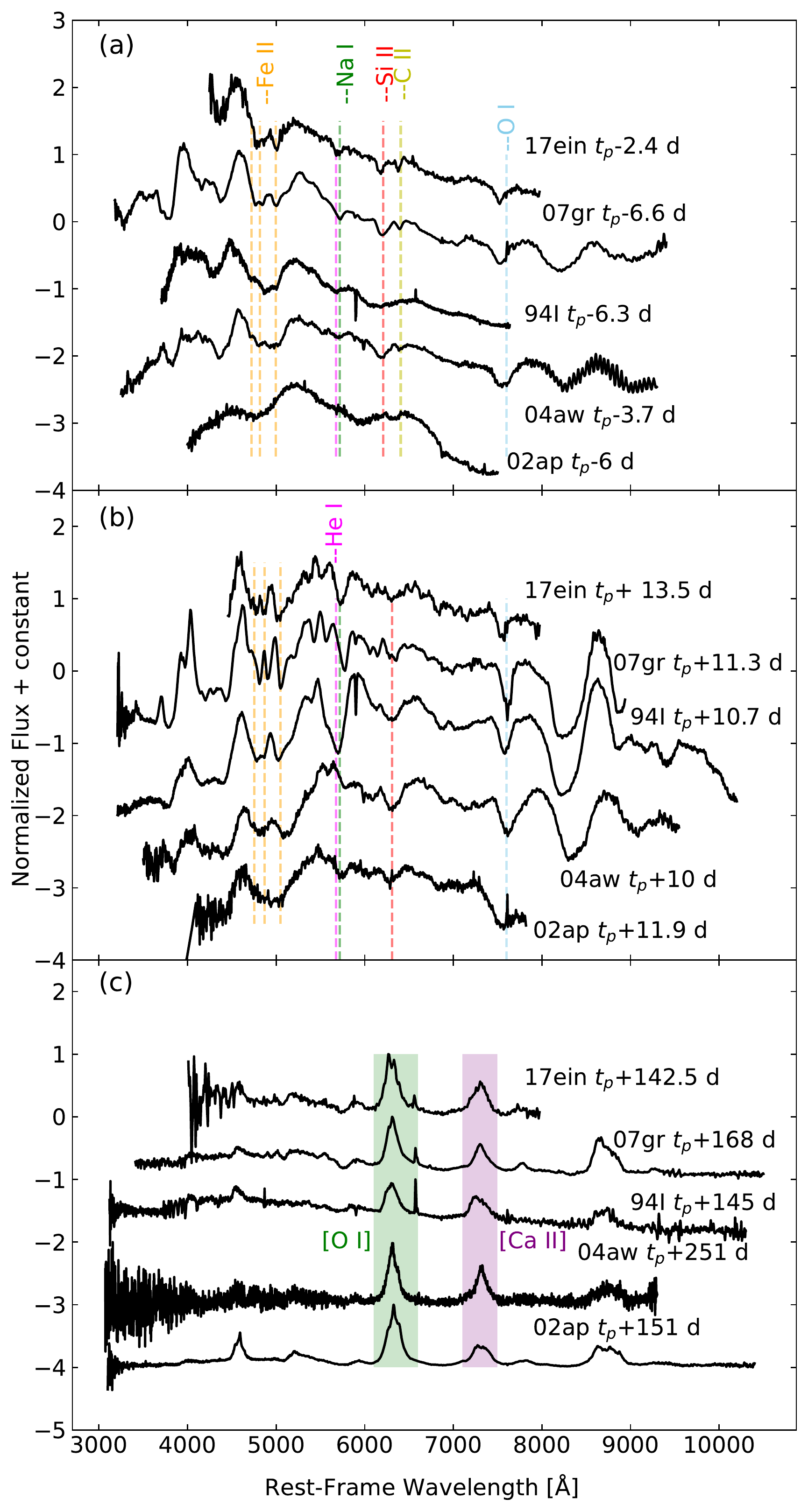}
	\caption{Selected epochs compared to a small sample of other Type Ic SNe (SN~1994I \citep{Filippenko1995}, SN~2007gr \citep{Valenti2008_07gr}, SN~2004aw \citep{Taubenberger2006}, SN~2002ap \citep{Foley2003}). These SNe are chosen as SN~1994I and SN~2004aw are both Ic-6 SNe, SN~2002ap is a Type Ic-BL or a Ic-4, and SN~2007gr is a close analogue to SN~2017ein and also a Ic-7. The \HeI\ is included as SN~1994I is thought to possibly contain He due to the stronger than normal feature.}
	\label{fig:similar_spec}
\end{figure}

Figure \ref{fig:similar_spec}(a) shows a small sample of Type Ic SNe compared to SN~2017ein at a epoch of approximately 1 week prior to bolometric peak. While these dates do not always reflect the epoch with respect to the explosion date, they offer a quick comparative time if the explosion date is not well constrained. SN~2017ein and SN~2007gr both share narrow features, with the \FeII\ \lam\lam4924, 5018, 5169 lines being separated and easily identifiable. These features are only partially separated in SN~1994I and SN~2004aw, while SN~2002ap does not show the same \FeII\ features. 
Figure \ref{fig:similar_spec}(b) shows the same SNe approximately 2 weeks after the time of the bolometric max, with SN~2007gr and SN~2017ein showing similar spectral behaviour over the shared wavelength range. SN~1994I and SN~2004aw also share similarities with each other, with the exception of the dominant feature near 5500 \AA, thought to be helium \citep{Filippenko1995,Sauer2006}. The nebular epochs in Figure \ref{fig:similar_spec}(c) show fewer differences between the set of SNe with again, SN~2007gr showing the most similarity to SN~2017ein. 


\section{Modelling} \label{sec:syn_mod}

We present the model for SN~2017ein following the abundance tomography method \citep{Stehle2005} used to model other SNe such as SN1994I \citep{Sauer2006} and SN~2004aw \citep{Mazzali2017}. The spectra cover a temporal range of $-12.5$ days to +192 days with respect to the pseudo-bolometric peak, with a well sampled photospheric phase that covers the --12.5 to +34.2 day range and 3 nebular spectra at +140 to +192 days. Of the photospheric spectra, we select 6 spectra in total, three pre-\tp, and three post-\tp, that show the spectral evolution and have a good S/N. 

We start with an initial density structure based upon a 22 \msun\ progenitor model. This model is stripped of all H/He to produce a CO core of approximately 3.4 \msun\ prior to core collapse.  \citet{Teffs2020} explored this model in detail as part of a parameter study that included four explosion energies of 1, 3, 5, and 8 foe, where 1 foe is defined as 1$\times$ $10^{51}$erg. This model had good success in reproducing a set of observed spectra from multiple Type Ic SNe without fine tuning any parameters, so we treat the density and abundance structure of the 1 foe model as our base. Given the estimated ejecta properties in \citet{VanDyk2018, Kilpatrick2018, Xiang2019} and the results from our parameter study from \citet{Teffs2020}, the initial CO core model is likely too massive to reproduce both the spectra and photometry due to a slower diffusion time post peak and the lack of similarity to SN~1994I shown in \citet{Teffs2020}. To find a better fit for SN~2017ein, we rescale the model using the method in \citet{Hachinger2009} by using equations \ref{eqn:dens_scale} and \ref{eqn:vel_scale}.

\begin{equation}
\label{eqn:dens_scale}
\rho^{'}  = \rho \Biggr( \frac{E_{k}^{'}}{E_{k}}\Biggr)^{-3/2} \Biggr(  \frac{M_\mathrm{ej}^{'}}{M_\mathrm{ej}} \Biggr)^{5/2}
\end{equation} 

\begin{equation}
\label{eqn:vel_scale}
v^{'}  = v \Biggr( \frac{E_{k}^{'}}{E_{k}}\Biggr)^{1/2} \Biggr(  \frac{M_\mathrm{ej}^{'}}{M_\mathrm{ej}} \Biggr)^{-1/2}
\end{equation}
The primed variables in equations \ref{eqn:dens_scale} and \ref{eqn:vel_scale} are the new, rescaled variables and un-primed are the original variables. The model is re--scaled to some initial parameters for \ek\ and \mej, with the initial \mni\ derived from the peak luminosity. The scaled density profile is then used in our Monte Carlo light curve code, which tracks the emission and propagation of $\gamma$-rays and positrons produced by the decay of \Nifs, and subsequently \Cofs, in to the homologously expanding ejecta as described in detail in \citet{Cappellaro1997}. The outputted bolometric light curve is then compared to the observed pseudo-bolometric light curve and the process is repeated until the calculated light curve reproduces the observed light curve with reasonable accuracy.

The re--scaled model and its initial abundances are then used in our spectral synthesis code, described in detail in \citet{Mazzali1993,Lucy1999,Mazzali2000b}. This code reads in the composition and the density profile of the ejecta to produce a stratified model where each layer is defined by an observed spectrum and fitted abundances, photospheric luminosity (\lph), and photospheric velocity (\vph). These parameters are systematically changed until the synthetic spectrum best reflects the observed spectrum and its behaviours. By using abundance tomography, this allows us to create a stratified ejecta that gives us both the abundance and distribution of elements responsible for the formation of the spectral features. For elements that do not produce strong optical features or that produce features beyond the observed spectral wavelength range of SPRAT (\lam\ > 8000 \AA), such as the strong \CaII\ feature near 9000 \AA, we are unable to constrain the abundance of those elements.

 \begin{table}
 	\centering
 	\caption{Synthetic model properties for SN~2017ein from the nebular and photospheric models. The error bounds in the \Nifs\ mass comes from the modelling of the nebular and photospheric phases, while the others are estimates from the light curve modelling and the parameter study of \citet{Ashall2020}.}
 	\begin{tabular}{cccc}
     \hline
 	Mass & \mni\, & \ek\,  & \eom\, \\
 	
 	[\msun] & [\msun] &  [10$^{51}$erg] & [10$^{51}$erg /\msun] \\

     \hline

 	1.6$\pm$0.4	 &	 0.09$\pm$.01	 &	0.9$\pm$0.2&	0.56$\pm$.05 \\
     \hline
 	\end{tabular}
 	\label{tab:mod_props}
 \end{table}
Using these methods, we find a model with an \mej\, of $1.6 \pm0.4$ \msun, $0.09\pm0.01$ \msun\ of \Nifs, and an \ek\ of $0.9\pm0.2$ foe that reproduces the bulk of the spectral features and the pseudo-bolometric light curves, with error estimates taken from the error analysis by \citet{Ashall2020} and shown for simplicity in Table \ref{tab:mod_props}. In order to reproduce the velocity of the \OI\ \lam 7774 line early in the spectra, the outermost region, or $v > 12000-27000$ \kms, is reduced in density by expanding the shell. This outer reduction in density may be a result of a low mass He shell not present in the initial model or some asymmetry.

\begin{figure}
	\centering
	\includegraphics[scale=.54]{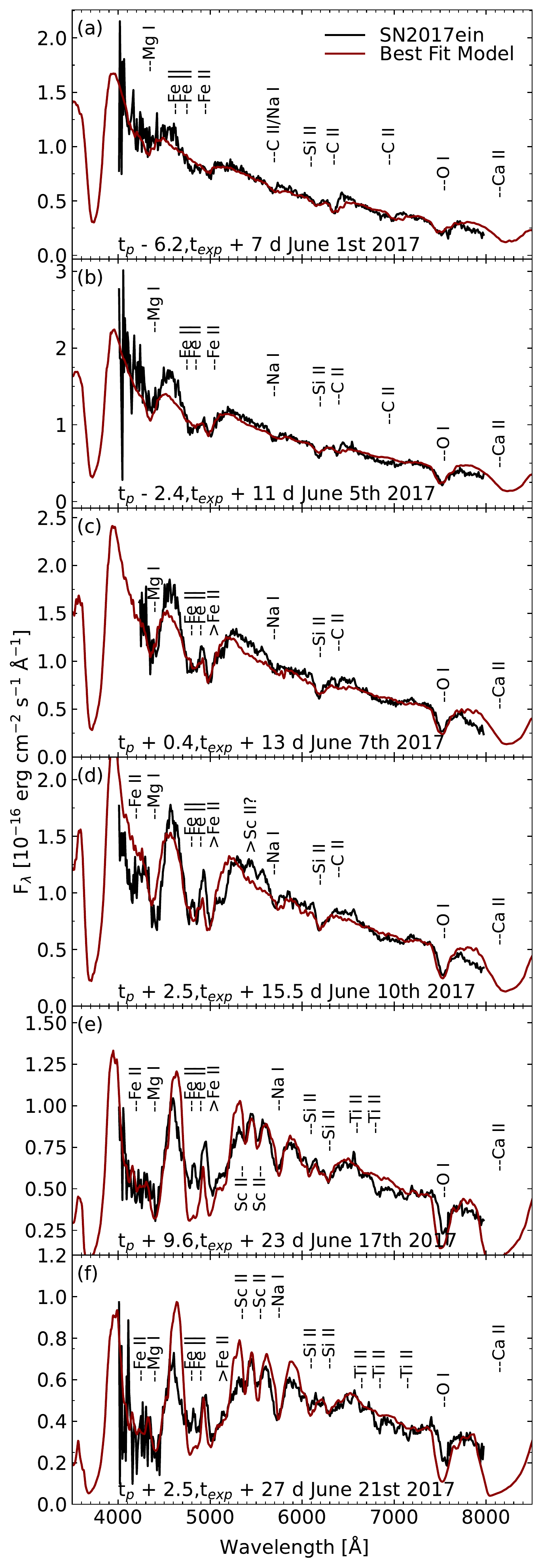}
	\caption{The spectral models for best fit model model for a selection of spectral epochs. Line identifications are for strong, dominant, or common features observed in SN~2017ein and other Type Ic SNe. The spectral models do contain other weaker lines that contribute to the spectral behaviour but are not listed.}
	\label{fig:models_nohe}
\end{figure}

\subsection{June 1st, 2017; \tp\ $-6.2$ d, \texp\ $+ 7$ d }\label{subsec:epoch1} 
The epoch shown in Fig.~\ref{fig:models_nohe}(a) is not the earliest available epoch, but shows good S/N with well defined features compared to the earlier spectrum. The best fit synthetic spectrum has a \vph\ of 11\,000 \kms\ and an Log$_{10}$(\lph) = 42.295 \ergs. The \vph\ and \lph\ parameters can be modified by of 5-10 $\%$ before the fit starts to visibly suffer, and is discussed in more detail in \citet{Ashall2020}. This early observed spectrum has only a few strong features. The re--emission red of the \CII\, line at 6500 \AA\ is not well reproduced in the model and the synthetic \CII\, lines are all slightly too broad. Despite this, the overall flux level and majority of features are well reproduced.

\subsection{June 5th 2017; \tp\ $- 2.4$ d, \texp\ $+ 11$ d }\label{subsec:epoch2} 
The best fit model spectrum in Fig. \ref{fig:models_nohe}(b) has a \vph\ of 9300 \kms\ and an Log$_{10}$(\lph) = 42.395 \ergs. Most features are well reproduced in the synthetic model, but the \CII\ features at 6500 \AA\ is too broad when compared to the observed spectrum. In addition, the flux level in the near--UV is slightly under the observed spectrum, but is necessary to reproduce the remaining features.

\subsection{June 7th 2017; \tp\ $- 0.4$ d, \texp\ $+ 13$ d }\label{subsec:epoch3}

The model spectrum in Fig. \ref{fig:models_nohe}(c) has a \vph\ of 8800 \kms\ and an Log$_{10}$(\lph) = 42.410 \ergs.  The \FeII\ lines are very well reproduced at this epoch, but the flux level is lower than the observed in the 5000--6500 \AA\ region. Similar to the previous epoch, more Fe--group material may improve this at the cost of the \FeII\ features. The \CII\ line at 6200 \AA\, is narrow in the observed spectrum, but too broad in the synthetic spectrum. There may be a feature near 5600 \AA\, but is hard to discern due to noise in the spectrum. This may be a \SI\, line or an early \ScII\, line but is not reproduced in our models.

\subsection{June 10th 2017; \tp\ $+ 2.5$ d, \texp\ $+ 15.5$ d }\label{subsec:epoch4}
The model spectrum in Fig. \ref{fig:models_nohe}(d) has a \vph\ of 8300 \kms\ and an Log$_{10}$(\lph) =  42.395 \ergs. The region between 5200--5800 \AA\ is poorly reproduced in this model despite the remaining features showing strong similarity. The next epoch shows these may be the beginnings of the \ScII\ features that become dominant in the next epoch. By this epoch, the \CII\ line is blended, too weak, or largely absent and the \NaI\ D line is beginning to become strong.

\subsection{June 17th 2017; \tp\ $+ 9.6$ d, \texp\ $+ 23$ d}\label{subsec:epoch5} 
\begin{figure*}
	\centering
	\includegraphics[scale=.7]{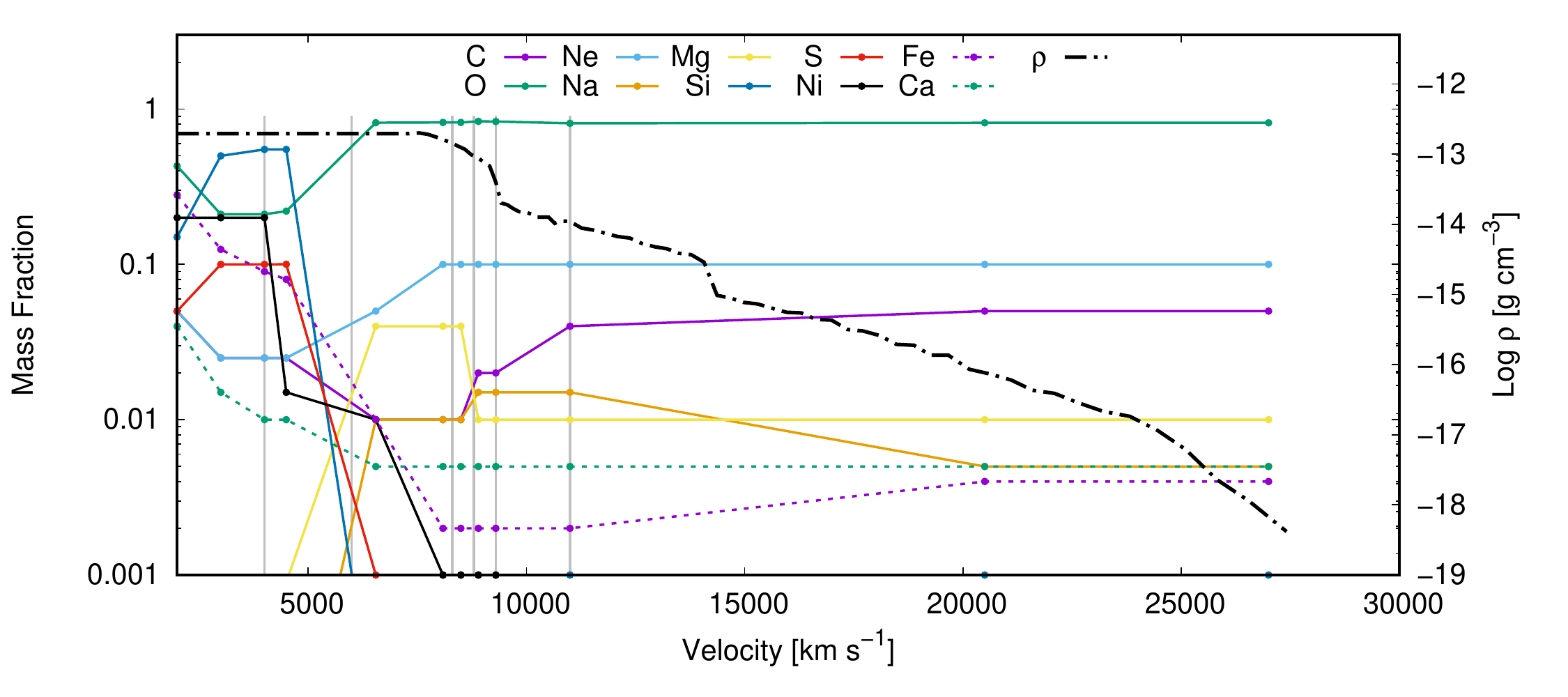}
	\caption{Mass fraction of the most abundant elements in the best fit model and the density profile (in black) shown on the right $y$-axis. The grey lines represent the modelled epochs with the \vph\ values given in Sections \ref{subsec:epoch1} to \ref{subsec:epoch6}.}
	\label{fig:abu_plot}
\end{figure*}

By this epoch, the observed spectrum is relatively low in flux, noisy, and has a multitude of weak features in the spectrum. The model spectrum in Fig. \ref{fig:models_nohe}(e) has a \vph\ of 6000 \kms\ and an Log$_{10}$(\lph) =  42.225 \ergs. The previously tentatively identified \ScII\ features are now strongly matching to the observed spectrum. The abundances of Sc and velocity range required to produce these lines are given in Figure \ref{fig:abu_plot}. Increasing the velocity range or mass of Sc can cause these lines to dominate in the later spectra. The \FeII\ feature is too broad and deep, but is required for the rest of the spectrum to match. The region between 6200--7300\AA\ has contributions from \OI, \FeII, and \TiII\ lines and is difficult to match each one as the features are not strong and the spectrum is noisy..

\subsection{June 21st 2017; \tp\ $+13.5$ d, \texp\ $+ 27$ d}\label{subsec:epoch6} 
For the final photospheric epoch modelled in this work, the synthetic model Fig. \ref{fig:models_nohe}(f) has a \vph\ of 4000 \kms\ and an Log$_{10}$(\lph) =  42.10 \ergs. At this phase, the photospheric velocity becomes harder to define as nebular emission, evident in the \CaII\,NIR triplet in the wider wavelength range spectra given in \citet{VanDyk2018,Kilpatrick2018,Xiang2019}, is a non-negligible contribution to the total observed flux. As the spectral synthesis code strictly treats the photospheric phase, this prevents a hard definition of the photospheric velocity, so the photospheric velocity value is the one required by the spectral shape and total flux. Beyond that, the rest of the spectrum is well reproduced. The \NaI\ D lines fit could be improved with the addition of more \FeII\ re--emission or a stronger \SI\ contribution.

\subsection{[\OI] Asymmetry in the Nebular Spectra}\label{subsec:OI_asymmetry}
\begin{figure}
	\centering
	\includegraphics[scale=.6]{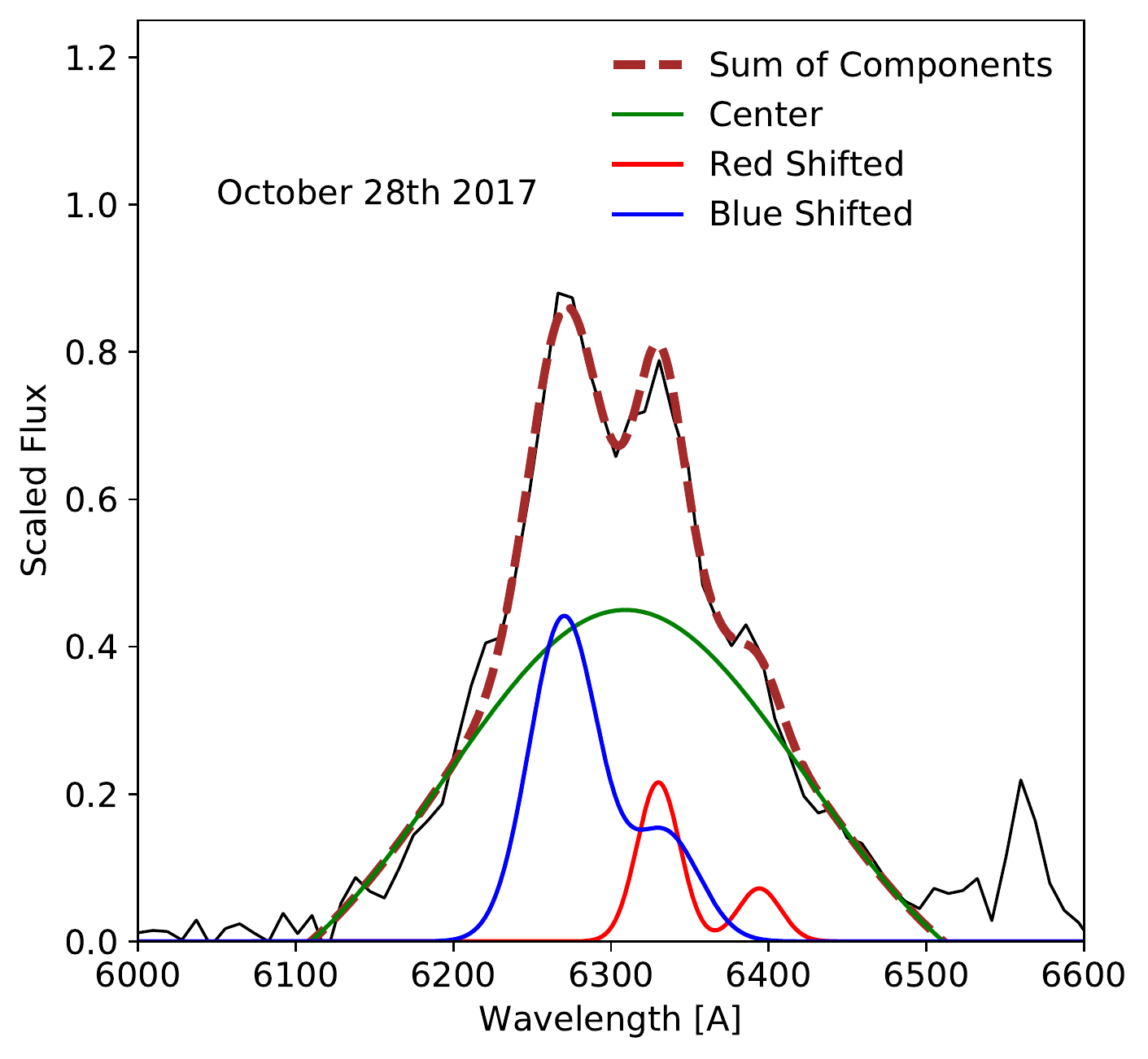}
	\caption{ Fits to the [\OI] \lam\lam 6300,6363 doublet for the October 28th 2017 spectrum of SN~2017ein using the same method as in \citet{Taubenberger2009}. The region around the doublet is normalised and scaled before the fit is applied.}
	\label{fig:taub_fit}
\end{figure}
The [\OI] \lam\lam 6300,6363 doublet is typically the strongest feature in the nebular spectra of Type Ib/c SNe and forms in a region free from other dominant features. The shape of this feature is often used as a probe for both the total oxygen mass of the innermost ejecta and as a way to estimate asphericity in these SNe. \citet{Taubenberger2009} modelled the emission line using a combination of up to 3 \Oneb\ components, each comprising of two Gaussian features that form a simplified [\OI] \lam\lam 6300,6363 doublet emission. 
Fits using up to three components reflect several possibly ejecta geometries, such as spherical or aspherical as well as torus--like structures. 

We apply this method on the October 28th spectrum in Fig. \ref{fig:taub_fit} and the following two epochs, but show only this epoch for simplicity as the results are similar. A similar [\OI] profile is shown for the three nebular spectra in Figure \ref{fig:spectra}. When compared to a set of other Type Ic in Figure \ref{fig:similar_spec}, SN~2007gr and the others do not show a similar [\OI] profile. The secondary bump near 6400 \AA\ and the approximately equal flux values of the two doublet peaks suggested two blobs of oxygen shifted by $\pm$30 \AA. Driven by this, we use three Gaussians comprised of a red and blue shifted blob and a central blended feature. We find a redshifted blob at a velocity shift of $-1200$ \kms\ and a blueshifted blob at a velocity shift of 2000 \kms. From \citet{Taubenberger2009}, a combination of these three components best reflects an asymmetric ejecta.

\subsection{Nebular Spectral Modelling}\label{subsec:nebular_model}

In order to establish the properties of the inner \mej, \Nifs\ mass and distribution in particular, we modelled the nebular-epoch spectra of SN\,2017ein. 
Three spectra are available, at rest-frame epochs of 160, 207 and 212 days from the putative time of explosion. They all show emission lines of the elements that are typically seen in SNe\,Ib/c. Unfortunately, the signal-to-noise ratio of the spectra is not always very high, especially in the blue. The region short of 5000\,\AA\ is extremely noisy, which makes it very difficult to use the strength of the [\FeII] lines as a test for the mass of \Nifs. Although this somewhat limits our models, we still use the overall flux to estimate the Fe flux. The strongest lines that are seen are [\OI]\,\lam\lam6300,6363 and \CaII]\,\lam\lam7291,7324. \MgI]\,\lam4570 may also be seen, while a broad emission near 5200\,\AA\ may be identified as a blend of [\FeII] lines, but it is affected by noise. 

A particularly interesting aspect of the spectra is the multi-peaked structure of the [\OI]\,\lam\lam6300,6363 line. This is not uncommon in SNe\,Ib/c and in some cases, very widely separated peaks or an abnormally narrow [\OI] line suggest a highly aspherical explosion. These are typically seen in energetic events like GRB/SNe \citep{Mazzali2001,Mazzali2005}, but these events are quite rare. More often narrowly separated peaks are seen \citep[e.g.,][]{Maeda2008,Taubenberger2009}, or composite profiles \citep[e.g.,][]{Mazzali2017}. These may be interpreted as asymmetries that affect only the innermost ejecta, which may carry the signature of an asymmetric or aspherical behaviour in the collapse/explosion, which was however not strong enough to affect the entire progenitor and then the SN ejecta as a whole. The rate of occurrence of this type of profiles is quite high, suggesting that most SNe\,Ib/c are affected to some degree by inner asphericity \citep{Maeda2008}. 
\begin{figure}
	\centering
	\includegraphics[scale=.4]{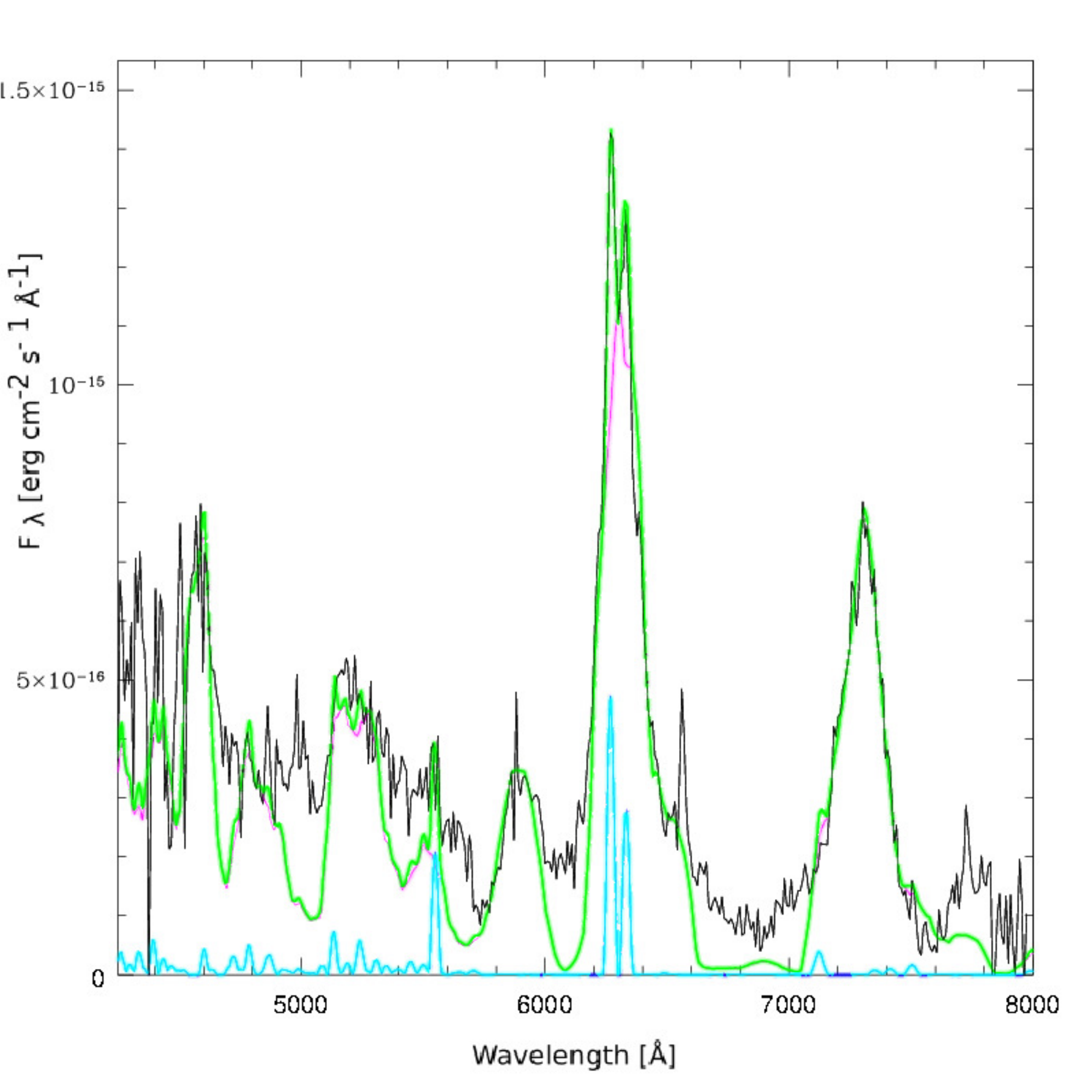}
	\caption{The nebular model for the observed spectrum, in black, at day 160, where the blue line represents the flux from a small mass of blueshifted oxygen, the pink from a more massive central mass of oxygen, and the green is the combined flux from both contributions}
	\label{fig:neb_day160}
\end{figure}

In the particular case of SN\,2017ein, however, the profile analysis above suggests the presence of components at different bulk velocities. This makes it interesting to model the spectra using a multi-component scenario as done for the peculiar SN\,Ia 2007on \citep{Mazzali2018}. 
For the modelling, we used our SN non-local thermodynamic equilibrium (NLTE) nebular spectral synthesis code. The code has been discussed and used in several papers, dealing with both SNe\,Ib/c \citep[e.g.,][]{Mazzali2007,Mazzali2017} and Ia \citep[e.g.,][]{Mazzali2008,Mazzali2020}. Given a density and composition, the code first computes the emission of gamma-rays and positrons from the decay chain \Nifs\
$\rightarrow$ \Cofs\ $\rightarrow$ \Fefs. 
This can be done in both a 1-zone or a stratified approach, where density and abundances change with radius in a one-dimensional setup. The propagation and deposition of the gamma-rays and positrons is then computed using a Monte Carlo method. Typical opacities that are used are $\kappa_{\gamma} = 0.027$\,cm$^2$\,g$^{-1}$ and $\kappa_{e^+} = 7$\,cm$^2$\,g$^{-1}$. 

The deposited energy heats and ionizes the gas and, following \citet{Axelrod1980}, ionization is assumed to be caused by impact with the high-energy particles produced in the deposition of gamma-ray and positron energy, while photoionization is assumed to be negligible \citep{Kozma1998}. Ionization balance is obtained matching impact ionization and the rate of recombination for each ion, which depends mostly on density. Level population is computed in NLTE, balancing the thermal heating rate with cooling via line emission. This takes place mostly via forbidden lines, but some permitted transitions that can be strong in emission and contribute therefore to cooling are also included, e.g.\ \CaII\ IR triplet and H\&K. The nebula is assumed to be optically thin, therefore radiation transport is not performed. This is a reasonable approximation at late times. Line emission then tracks the distribution of the emitting elements and that of \Nifs. 

Given the profile of the [\OI] line, it is clear that a single, spherically symmetric emission component would not be able to reproduce the observations in detail. Therefore we take a different approach. Using the abundance distribution derived from the early-time spectra, we model the broad component of the emission, completing the mapping of the density and abundances at low velocities as required to optimise the fit. We then add a narrow, blue-shifted component to reproduce the narrow peaks in the observed [\OI] line. Finally, we test our results by evolving the model to different epochs and comparing to the observations. We note a discontinuity in the light curve when one compares points near 150 days and 200 days (Fig. \ref{fig:lc}). This poses a problem for both nebular and light curve modelling, which are based on the same treatment of gamma-ray and positron deposition. 

As a first step, we determine the width of the [\OI] \lam\lam6300,6363 emission using a 1-zone approximation. A value $v = 5000$\,\kms\ is found to be appropriate at all epochs, when line blending is accounted for. This confirms that most \Nifs, as well as a significant fraction of oxygen, must be located inside this velocity. The broad emission profile is reasonably symmetric, suggesting that a 1D approach is sufficient for the bulk of the inner ejecta.  Using the model above, which has \mej\ $\approx$ 1.6\,\msun\ and \ek\ $\approx 9.0 \times 10^{50}$\,erg\,s$^{-1}$, we find that a \Nifs\ mass of $\approx 0.078$\,\msun, combined with an oxygen mass of $\approx 0.86$\,\msun\ yields a good fit to the broad feature at day 160. 

When including the other elements at low velocity, we need to remember that we have no direct handle on important elements such as silicon and sulphur, whose strongest emission lines are predominantly in the near-infrared (NIR). Depending on the amount of these elements that is included in the model, the overall mass can change significantly \citep[e.g.,][]{Mazzali2019}. Here we chose to keep the total mass as defined by the early-time modelling, which limits the mass of Si to $\approx 0.60$\,\msun\ and that of S to $\approx 0.07$\,\msun, most of which is at velocities between 3000 and 7000\,\kms. Calcium is responsible for the second strongest observed emission line, \CaII]\,7291,7324. The strength of this semi-forbidden transition allows us to determine the local density, which we achieve by setting some degree of clumping in the ejecta.  We use a volume filling factor of 0.1 for the inner ejecta, which is in line with previous results for SNe\,Ib/c \citep[e.g.,][]{Mazzali2010}, and a total Ca mass of $\approx 0.014$\,\msun. The synthetic spectrum obtained for day 160 is shown in Fig. \ref{fig:neb_day160}.

A narrow-line emission spectrum is then computed as a 1-zone model, and added to the broad-lined spectrum. A 1-zone approach is appropriate because, given the resolution of the observed spectra, the narrow emission profiles are only approximately described. Typically, the narrow [\OI] emission has a velocity width of $\approx 1100$\,\kms, which means that the two components are individually observed. The fact that they are in a ratio of about 2/3 implies that the local density is quite high, so again we need to use a volume filling factor of 0.1 for this clump of material. We assume that this clump is mostly composed of oxygen and heated by \Nifs\ decay. This is obviously an over-simplification, as given its low blueshift the clump is likely to be residing inside the bulk of the ejecta and therefore to be exposed to radioactivity from the rest of the \Nifs. In any case, under the approximation we used, it is sufficient to place 0.10\,\msun\ of oxygen in a clump heated by $0.023$\,\msun\ of \Nifs\ to reproduce the observed narrow emission. When added to the broad component, a reasonable fit is obtained for the whole profile (Fig. \ref{fig:neb_day160}). 

The profile decomposition performed above suggested the need for three components, but we do not need that when we compute our synthetic models: the two main narrow peaks correspond to [\OI]\,6300 and 6363, respectively, not to a blue-and red-shifted narrow component. A 3-component model may explain the small ledge near 6400\,\AA\ as red-shifted [\OI]\,6363, but it fails to explain the corresponding ledge near 6200\,\AA. It is actually more likely that these two ledges, if they are indeed real and not just due to noise, are features of the overall density/abundance distribution, as they are rather symmetric with respect to [\OI]\,6300, at velocities of $\sim 4000$\,\kms. The narrow-lined spectrum shows emission also in [\OI]\,5577, which is a tracer of recombination, and which may correspond to an observed feature. It also shows several [\FeII] lines, but these are too weak to be distinguished in the noisy observed spectra. 

If we evolve the model to later epochs,  we can only obtain a good fit to the observations if we apply multiplicative correction factors of 1.3 for the spectrum on day 207 and 1.6 for that on day 212. Such rapid changes in the luminosity are unlikely unless we invoke improbable sudden changes in the gamma-ray or positron opacities. We discuss this issue in further detail in Section \ref{subsec:late_time}.

In conclusion, the nebular models largely confirm that \Nifs\ is concentrated to low velocities, as is expected in a low-energy explosion. The mass of \Nifs\ that is derived, $\sim 0.08$\,\msun, is consistent with that obtained from the light curve. An interesting question is the origin of the oxygen-rich blue-shifted clump. We do not see evidence of a counter-clump at some red-shift. It is possible, but unlikely given the symmetric profiles of the broad emission features, that such a clump might exist but its emission is absorbed by the bulk of the ejecta. The innermost layers of the ejecta may be characterised by some degree of asphericity, which may have been imprinted upon them at the time of the explosion. The mass of the clump is however quite small, and it would be even smaller if we actually embedded it in the \Nifs-dominated inner ejecta. It does not seem too likely, given the small amount of material comprising the clump, that the asphericity is caused by jets, but if that is indeed the case the jet(s) were weak and easily choked inside the dense deep CO core of the star. Such events may indeed be quite common in SNe\,Ib/c \citep{Piran2019}, and possibly in all core-collapse SNe \citep{Nakar2017,Gottlieb2021}

\subsection{Light Curve}\label{subsec:fit_lc}
\begin{figure}
	\centering
	\includegraphics[scale=.49]{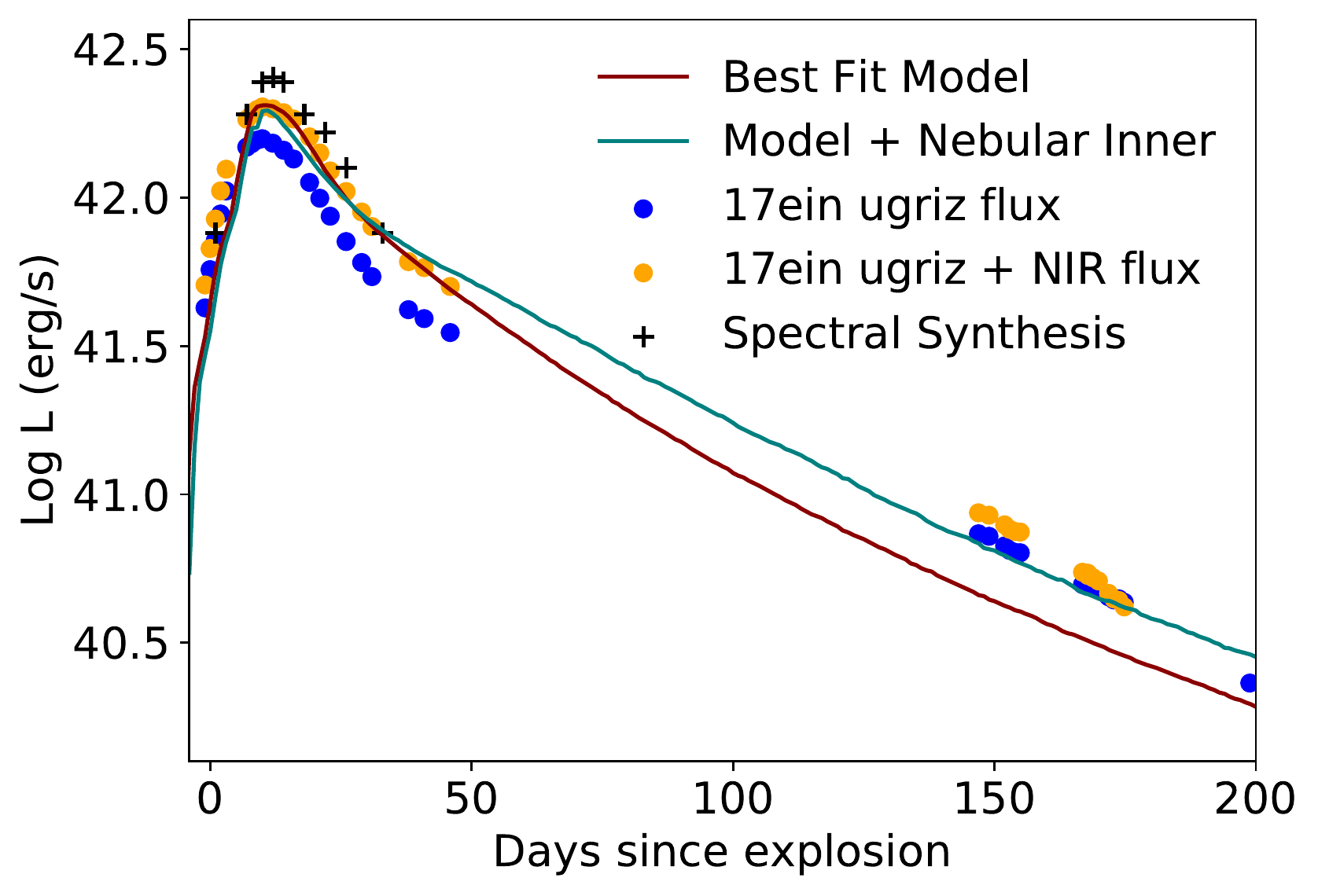}
	\caption{Bolometric light curves for both SN~2017ein, two synthetic models, and the integrated flux from the spectral models }
	\label{fig:bolo_lc_fits}
\end{figure}

We compute a synthetic light curve using a Monte Carlo code discussed in detail in  \citet{Cappellaro1997} and \citet{Mazzali2000b}. The code requires a choice of density structure and elemental abundances including \Nifs.  The code tracks the emission and propagation of $\gamma$-rays and positrons produced by the decay of \Nifs, and subsequently \Cofs, into the expanding ejecta. The gamma ray and positron opacities are given by $\kappa_{\gamma} = 0.027$\,cm$^2$\,g$^{-1}$ and $\kappa_{e^+} = 7$\,cm$^2$\,g$^{-1}$ \citep{Axelrod1980}. The deposited energy from the decay is recycled into optical photons and the resulting propagation is followed using a similar Monte Carlo scheme. The optical opacity is both time and metallicity dependent, which aims to reproduce the dominance of line opacity in the ejecta \citep{Ashall2019,Mazzali2000b}. This code is used initially to produce an approximate model for the model inputs in the spectral synthesis code. After the photospheric and nebular spectra are fit, any changes to the model in density or abundances are used to produce a final light curve shown in Fig. \ref{fig:bolo_lc_fits}.

The mass of \Nifs\ required to match the flux level of the nebular spectra is approximately 0.08 \msun\ compared to the 0.1 \msun\ of \Nifs\ required to match the pseudo-bolometric light curve in Figure \ref{fig:CompareBolometrics}. Photometric observations for SN~2017ein cover only the $ugriz$ bands which cover a wavelength range of 3000 -- 10000 \AA, with no UV or IR contributions. \citet{Sauer2006} used the NIR/IR flux from SN2002ap, scaled to SN1994I to calculate an estimated NIR ($\lambda$ > 10000 \AA) flux to generate a more complete bolometric light curve. SN2002ap was chosen due to being a Type Ic SNe with well sampled NIR photometry. For SN~2017ein, both its spectral evolution and light curve behaviour is similar to that of SN~2007gr \citep{Valenti2008_07gr,Hunter2009}, and has $JHK$ band (10000 -- 25000 \AA\,) coverage for both near peak and late time epochs. We follow the same method described in \citet{Sauer2006} to generate a modified bolometric light curve.

While the explosion models match the observed $ugriz$+NIR light curve, the bolometric light curve from our synthetic spectral modelling is too luminous by 5-10\%. This mismatch is likely the missing UV flux not taken into account in addition to the standard errors inherent to the observations and the modelling process that is discussed later. The modified inner density and \Nifs\, distribution required to match the nebular spectra is used to generate a second model. The modified model has a bolometric light curve that reproduces both early phase and late time phases fairly well. As only the innermost ejecta is modified, the spectra in Fig. \ref{fig:models_nohe} are unchanged.


\section{Discussion}\label{sec:discus}

\subsection{Model Properties}\label{subsec:mod_disc}

The initial model in this work is based upon the CO core of a 22 \msun\ progenitor model evolved at solar metallicity, with no binary, then stripped of all helium and exploded at 1 foe. The initial model with no modifications failed to reproduce the spectra and the light curve as did the higher \ek\ models in \citet{Teffs2020} using the same initial progenitor. This required us to modify the CO core model in order to best fit the observed spectra and light curve to get the results in Section \ref{sec:syn_mod}.

Based upon the estimates given in \citet{VanDyk2018, Kilpatrick2018, Xiang2019} and the similarity to SN~2007gr in Fig. \ref{fig:similar_spec}, we re--scaled the mass of the model to get a \mej\, of 1.6 \msun\, and the energy to match the relatively low velocity lines. We do note that re--scaling a model in either or both \ek\, and \mej\, may not reflect the properties of an evolved CO core of the re--scaled mass. Based upon only the photospheric modelling, we suggest that the \mej\, of SN~2017ein is within the range of 1.2--2.0 \msun, using the error estimates from \citet{Ashall2020}. The \ek\ was modelled to be $\sim$0.9 foe. These values place the \eom\, ratio to be $\sim$0.56  foe/\msun\, for the best fit model. Comparing these ratios to those found in \citet{Prentice2017}, the \eom\ ratio for our model is on the low end. Both SN~1994I and SN~2004aw have \eom\, ratios near 1 foe/\msun, but show broader spectral features than SN~2017ein or SN~2007gr. Using the classification system from \citet{Prentice2017}, both SN~1994I and SN~2004aw are defined as Type Ic-6 due to the blending of two of the three primary \FeII\, lines while SN~2017ein is a Type Ic-7. For modelled SNe, as the classification increases from very blended Type Ic-3 to narrow lined Type Ic-7, the \eom\, ratio decreases. Based on this, the low \eom\, ratio found for SN~2017ein, which is the first SN Ic-7 to undergo this kind of analysis, fits in with the previous trend. The modelling of the nebular phase spectra also matches the model from the photospheric phase.

\subsubsection{Scandium}\label{subsubsec:scandium_det}

From day 15 and onward, the region near 5500\AA\, contains two features between the strong \FeII\ and \NaI\ lines. A similar pair of lines were seen in SN~2007gr and \citet{Valenti2008_07gr} suggested these are \ScII\, lines but no abundance tomography model has been done for that event. \ScII\ lines have been in observed in Type II SNe, such as SN~1987a \citep{Mazzali1992} and the line velocities of the 5527 and 6245 \AA\, \ScII\ lines were measured in the Type II SNe 2012A, 1999em, 2005cs and 2009bw, summarised in \citet{Tomasella2013} \citet{Millard1999} tentatively identified \ScII\ in several spectra of SN1994I as well but \citet{Sauer2006} were unable to reproduce the identified \ScII\ line without producing other unwanted spectral lines.

For the models in Fig. \ref{fig:models_nohe}, a small mass of Sc is located between 8300 -- 11000 \kms. These two features are best fit in the June 17th or June 21st spectra, (Fig. \ref{fig:models_nohe}(e-f)). The addition of \ScII\ produces a set of lines between 6000--7000 \AA, but this is also combined with \FeII, \SiII, and various other weaker features. The June 10th spectrum shows what may be the beginnings of the \ScII\ lines, but the synthetic model has a weaker flux in this region. Increasing the \FeII\ mass can cause re--emission of flux into this area, improving the flux level but can produce unwanted behaviour in other regions of the spectrum or for later spectra. The earlier spectra do not show strong evidence for \ScII\ and the flux at the later epochs is fairly low. Increasing the mass of \ScII\ at a higher velocity can reproduce both the two lines and re-emits flux towards the \NaI\ D region, reducing the depth of that feature but producing much stronger features in this epoch and later epochs between 6000--7000 \AA\ that are not observed. Sulphur can also produce the redder feature, but given the constraints on S/Si ratio from explosive nucleosynthesis (1/3 -- 1/5), increasing the amount of sulphur to produce this line requires too much Si in the outer region. One can also use Cr to reproduce part of the bluer line, but the mass and location of the material needed in the model produces too many unobserved lines if the feature is solely formed from Cr, in addition to the lack of a physical origin for such high amounts of Cr.

\subsubsection{Helium}\label{subsubsec:he_detect}

Similar to other Type Ic SNe, SN~2017ein has been suggested to have some amount of He in the outer atmosphere. The synthetic model in Fig. \ref{fig:models_nohe} contains no He, and only Na is required to produce the feature at 5875\AA\, with good success throughout all the synthetic spectra. The two weaker optical \HeI\ lines at 6678 and 7065 \AA\ are not present or easily identifiable in any of the spectra and we have no NIR spectra to cover the stronger 10830 and 20581 \AA\ lines. Due to these factors, we would have to solely rely on the \HeI\ \lam5875 line to determine if He is present in the model. As this feature is dominant in the same region in which the \NaI\ D line is located, this is difficult. \citet{Hachinger2012} and \citet{Teffs2020b} found that up to 0.06--0.14 \msun\ of He could be present before the optical and NIR \HeI\ lines are strong and easily identifiable, with a much lower amount of ``hideable" He if only NIR lines are considered. 

\begin{figure}
	\centering
	\includegraphics[scale=.55]{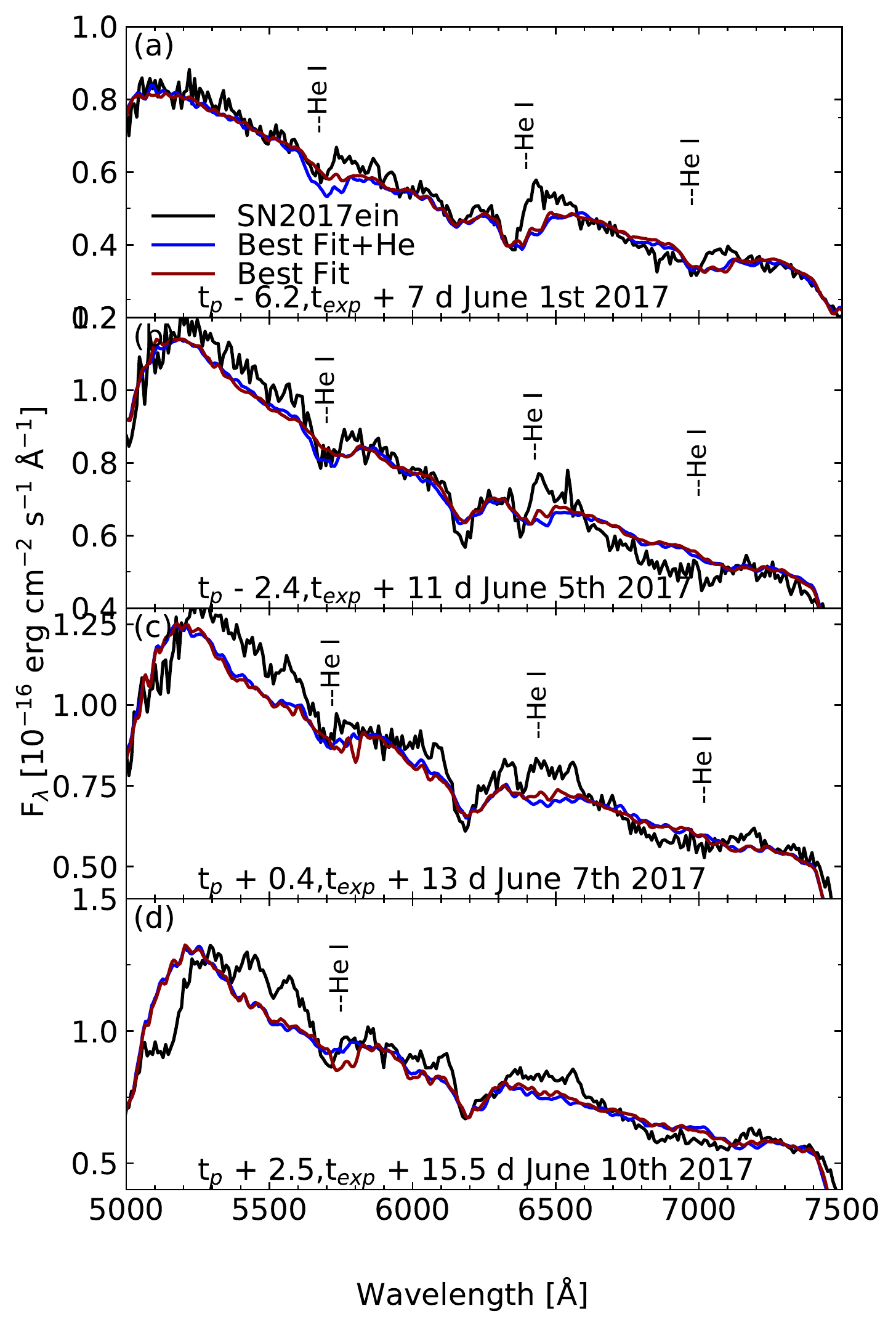}
	\caption{The best fit model with Na and no He and an alternate model with He and no Na for the first four epochs. The spectral range is constrained to 5000--7500 \AA\ in order to cover the three optical \HeI\ lines at 5875, 6678, and 7065 \AA. }
	\label{fig:early_epoch_he}
\end{figure}

Figure \ref{fig:early_epoch_he} shows the four earliest epochs with a total \mhe\ $\sim$0.01 \msun\ above a \vph\ $>$ 9300 \kms. The addition of He to this model (blue line in Fig. \ref{fig:early_epoch_he}) reproduces the primary optical \HeI\ line at 5875 \AA\ to an adequate degree, but the lack of the weaker \HeI\ lines at 6678 and 7065 \AA\ makes a single line fit hard to interpret as the model without He also matches the features adequately.  Increasing the amount of He beyond the $\sim$0.01 \msun\ can drastically strengthen the optical features beyond what is seen in the observed spectra. For SN~1994I, in which the 5875 \AA\ region shows a stronger feature than in other Type Ic SNe, the possible signature of \HeI\ is relatively weak and short lived. Models for SN~1994I with and without He can often reproduce the bulk behaviour, suggesting that if He is present in SN~1994I, and by relation SN~2017ein, the contribution from He would be minimal, especially compared to He dominated features in Type Ib SNe. As concluded in \citet{Hachinger2012} and \citet{Teffs2020b}, a combination of optical and NIR \HeI\ lines are needed to determine how much He is present in Type Ic SNe.

\subsection{Progenitor Implications}\label{subsec:prog_impli}

\begin{figure}

	\includegraphics[scale=.47]{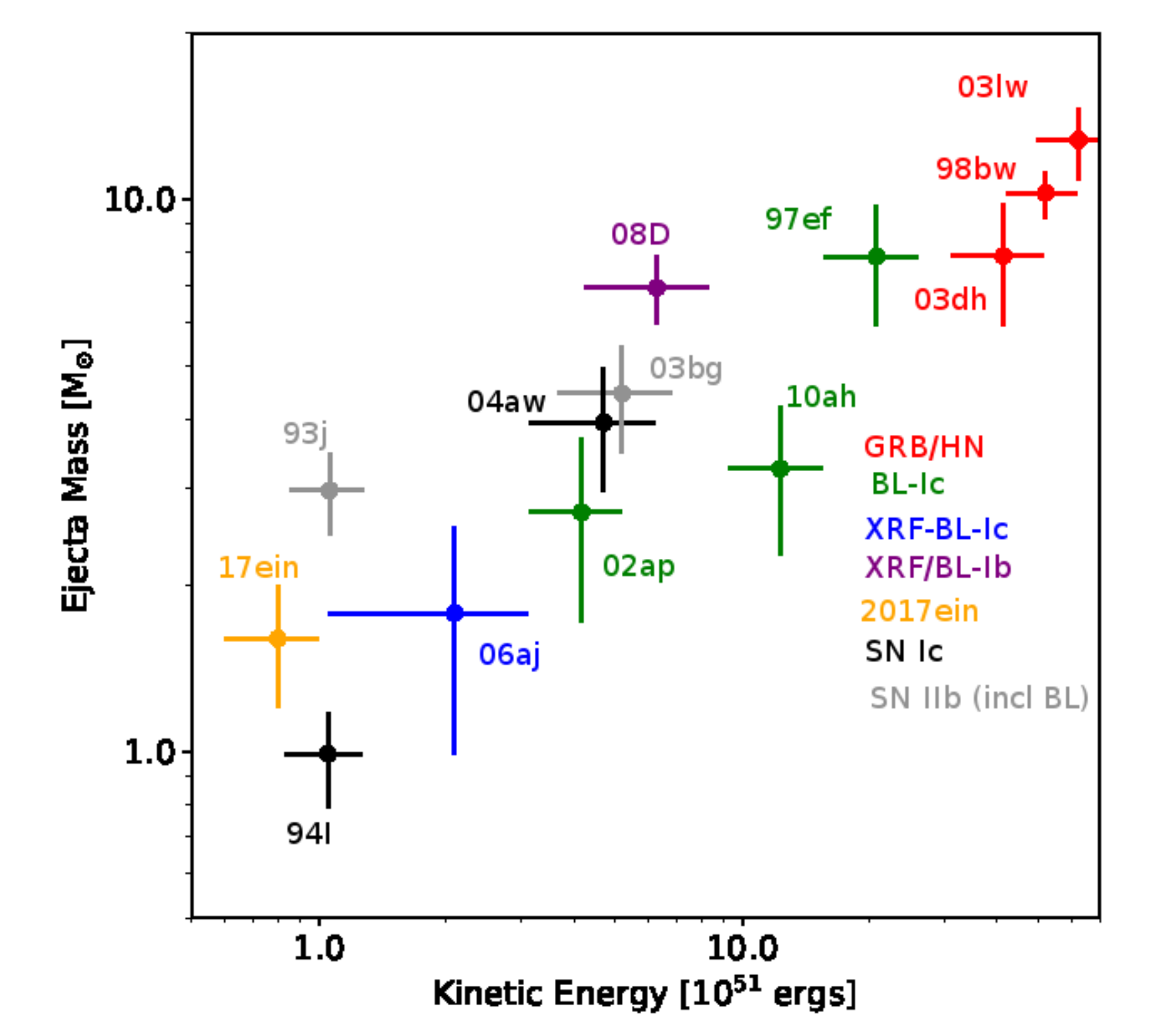}
	\caption{A sample of modelled results from a variety of SN types that show the explosion energy and \mej. See \citet{Mazzali2017} for individual SN references }
	\label{fig:kin_mej}
\end{figure}

The modelled values for the ejecta mass, explosion energy, and \Nifs\ mass in this work are similar to the approximate fits from either the pseudo-bolometric light curve or the multi-band photometry, in \citet{VanDyk2018}.  However, the estimated \mej\, from the progenitor masses found in \citet{VanDyk2018, Kilpatrick2018, Xiang2019} show a significant disparity. The ejecta masses from both the single star or binary star masses are in the range of 4-8 \msun. 
Figure \ref{fig:kin_mej}, derived from spectral models and summarised in more detail in \citet{Mazzali2017}, shows that a correlation exists between \mej\ and \ek. 
Placing a SN with 4-8 \mej\ and \ek\, $<\sim1 \times 10^{51}$ erg in this plane 
would result in a significant outlier compared with other modelled SN~Ic. 
Taking that same \ek\ and the $\sim2$ \msun\ ejecta mass found in this work, gives a point that does align with other objects.

The modelling of SN~2017ein in this work uses the abundance tomography method that has been used often for a wide variety of SE-SNe. This is a multi-variable approach to modelling SNe, and combined with photometric and nebular fits can produce a consistent model that reproduces both the photometry and the early and late time spectra within the limitations of the modelling framework. Recent work by \citet{Ashall2020} focused on how changes in the model parameters, such as \vph, \lph, and \mej\ can alter a fitted synthetic spectrum and found that modelling both the photospheric and nebular phase spectra results in approximately 5--10\% error bars for both the \ek\, and \mej. They also suggested that a well fitted spectrum seems to prefer a single combination of \lph\, and \vph\, with a synthetic spectrum deviating from the fit with a 5-10\% change in either or both the \lph\, and \vph.

Using the modelled ejecta mass of 1.5 $\pm$ 0.4 \msun, we can first consider the low mass (i.e. \mzams\, $\leq$ 22 \msun) models from \citet{Woosley_Langer1993, Woosley_Langer1995, Woosley1995}. The \mzams\, models that best reproduce a 1.2--2.1 \msun\, CO core are those in the 16--20 \msun\, range. These models are all a result of single star evolution. For the same model set, a \mzams\, $>$ 35 \msun\, Wolf--Rayet stage star has mass loss that produces CO cores with a narrow final mass range of $\sim$2--4 \msun. \citet{Pols2002} find a much higher CO core mass for the same mass range, as the resulting CO core masses are highly dependent on the Wolf--Rayet stage mass loss rates used \citep{Yoon2017}.

If we consider that the progenitor is a He star that has evolved in a binary, such as those generated by \citet{Woosley2019} and exploded in \citet{Ertl2020}, we can find mass cases where its possible that the resulting He envelope can be lost. In particular, He stars of masses 3-5 \msun, reflecting a \mzams\ of 15-20 \msun, were found to have either a rapid He envelope expansion that could transfer mass or depending on the choice of mass loss rates, could lose their He envelopes through stellar winds. The CO core masses for this range (1.3-2.7 \msun) also matches the model derived in this work.  The explosion properties of these models in \citet{Ertl2020} are within the range of our modelled \ek\ and synthesised \mni\ as well, but their \ek\ and \mni\ relation is not well constrained due to approximations taken in the production of \Nifs. In addition, ``off the shelf" SNe models often do not replicate the spectra and photometry of observed SNe, so additional care should be taken.

As mentioned previously, the progenitor properties determined by \citet{VanDyk2018, Kilpatrick2018, Xiang2019} suggests a massive star with an \mzams\, $\sim$ 60-80 \msun, that with reasonable assumptions results in a 4-8 \msun\ ejecta mass. This is at odds with the approximate modelling methods used in those works that find an \mej\, closer to 1-2 \msun. In addition, our modelling also favours a low \mej\ of 1.6 $\pm$ 0.4 \msun\, as well. This is not the first time a contradiction between an estimated progenitor mass derived from pre--supernova imaging and modelling as arisen.  The nebular phase of SN~2007gr was modelled by \citet{Mazzali2009} to have $\sim$1 \msun\ of ejecta below 5000 \kms\ with an estimated explosion energy of 1 foe. Other estimates for SN~2007gr give an \mej\ of 2--3.5 \msun\ by \citet{Valenti2008_07gr} and $\sim$1.2 \msun\ from \citet{Drout2011}, both with a similar energy range of $\sim$1 foe. These ejecta estimates favour a low mass progenitor, around 15-18 \msun\ as well. \citet{Maund2016} estimates a much higher progenitor mass of 40 \msun\ by associating the progenitor with nearby stellar populations although with SN~2007gr, no object was detected in the pre--supernova images. 

The pre--supernova imaging is limited to two Hubble images and \citet{Kilpatrick2018} and \citet{Xiang2019} show a slight offset in the position of the compact source and the supernova. While this offset is well within the error bounds in the methods used, a combination of this and the contradictory modelling results may suggest that the progenitor is instead a faint star within the line of sight of the compact source. Follow up Hubble imaging by other groups is planned and would determine if the compact source remains after SN~2017ein dims.

An alternative explanation is to assume that the progenitor star is a massive star or in a massive binary that matches the pre--supernova imaging and is stripped of all H/He resulting in 4--8 \msun\ of ejecta at the point of collapse. At collapse, a significant fraction of this material forms a massive remnant with the explosion still ejecting 1-2 \msun\ of material. This would allow both the modelling in this work and the pre--supernova imaging to be correct. The resulting remnant mass would be in the range of 2--6 \msun, which is in the range of observed black hole x-ray transients \citep{Wiktorowicz2014} and within some theoretical ranges for massive stars, depending on mass loss and binary/single star evolution \citep{Fryer2012}. \citet{VanDyk2018, Kilpatrick2018, Xiang2019} all consider both binary/single star evolution as a possible candidate. This explanation is partially fine tuned, as it arbitrarily requires a particular black hole mass to form at collapse for both results to work together.

\subsection{\Nifs\ Mass} \label{subsec:nic_mass}
Approximately $\sim$0.1 \msun\ of \Nifs\ was required in the photospheric models with 0.08 \msun\ required in the nebular model. Given the estimated error in both the observational data and the modelling, a difference of 0.02 \msun\ is reasonable. Other SNe modelled using a combination of both photospheric and spectral methods show a soft correlation between \Nifs\ mass and \ek\ with lower \Nifs\ masses favouring lower ejecta masses \citep{Mazzali2017}. 
\Nifs\ masses found from photometric methods\footnote{Note, it is currently suspected that the use of `Arnett's rule' for SE-SNe may lead to a $\sim50$\% overestimate in \mni\ \citep[e.g.,][]{Woosley2020}.} typically find similar results \citep{Drout2011, Lyman2016,Prentice2019}. The most notable of these is SN~1994I, discussed previously, which is estimated to have a similar \mej\ of 1-2 \msun, an \ek\ of $\sim$1 foe, and \Nifs\ mass of $\sim$0.07 \msun, with some dependence on the reddening assumed. 

If we compare the photometry and spectra of SN~2007gr to that of SN~2017ein, we find similarities at both early and late epochs, excluding the latetime drop in the $r$-band. \citet{Mazzali2009} used the nebular phase spectra of SN~2007gr and estimated a total \mej\ of 1--1.5 \msun. Several other estimates for SN~2007gr are summarised in \citet{VanDyk2018} but \citet{Drout2011} uses a multiband fit and gets a similar mass and an \ek\, < 1 foe which results in an \eom\ ratio less than 1, similar to what we have found for SN~2017ein, which explains the similarity in spectra. This similarity allows us to approximate the NIR contribution of SN~2017ein in the total observed flux. The addition of this flux increases the peak luminosity by approximately 10--20\%. This requires a slight increase in \Nifs\ mass in the explosion model, from 0.08 to 0.1 \msun, slightly different from the nebular model.  The light curve from the synthetic model was still too bright, but falls within the observational error estimates and the 5-10\% missing UV flux not considered in the other light curves.

\subsection{Late Time Nebular and Photometry}\label{subsec:late_time}
We note in Section \ref{sec:lc} and in Fig \ref{fig:rbands} that the late time photometry at $\sim$200 days after \texp\ shows a larger magnitude drop in the  $r$-, $i$-, and $z$-bands post 150 days, than in the $g$-band. This additional drop results in the bolometric light curve not following a linear decay as seen in other \Nifs\ powered SNe (Fig.~\ref{fig:CompareBolometrics}).  

As mentioned in Section \ref{subsec:nebular_model}, the first nebular spectrum model can be powered by a similar mass of \Nifs\ to that of the early light curve, but the next two phases show a drop in flux that is inconsistent with the initial model unless a multiplicative factor is included to the total flux.  In the next few subsections we discuss possible explanations to explain this drop.

\subsubsection{Fallback}\label{subsubsec:fallback}
The reduction of overall flux seen in the nebular model can be explained by removing mass from the model between the first and final nebular phase epochs. This reduction in mass results in less material being heated and therefore less overall flux for the entire wavelength range. Using accretion as a primary energy source is often used to explain super luminous events with broad early light curves, of which SN\,2017ein has neither. If we consider the semi--analytic model for accretion powered SNe found in \citet{Dexter2013}, we can assign the luminosity from accretion to only an arbitrary fraction of the total luminosity, with the majority of the luminosity from a \Nifs\, decay model. A fit in this manner would see any fallback that occurs soon after the explosion be of low mass, low accretion rate, or have a very low efficiency of energy transfer; all of which would only be needed to add to the total luminosity 100 days after peak. Given the 100 days of missing data, a fit of this nature is arbitrary and fine tuned. 

The other alternative is that the fallback only occurs at some late time (t $>$ 50 days). This may be induced through late time CSM interaction which produces a strong enough reverse shock to cause the innermost material to fallback to the remnant \citep{Dexter2013} or some other dense enough material. If this did occur, the material would have had to been ejected many years prior to the collapse for it to be at such a distance from the star. However, there is no evidence of any CSM interaction in the spectra nor the light curve, unless it occurred in the 100 days when the event was behind the sun, but this would be an unlikely convenience.

\subsubsection{Secondary Energy Sources}\label{subsubsec:interact}
As mentioned in Section \ref{subsubsec:fallback}, late time CSM interaction could generate fallback and is thought to be a common energy source for superluminous SNe or those with long lasting light curves. The typical conditions used to suggest CSM interaction are not found in either the light curve or the spectra. Given the drop in bolometric flux at day 150, the interaction phase would need to occur and stop during the 100 days the SN was not observed, and the drop we see is the fall of the light curve towards a \Nifs\ decay curve.

A magnetar as a primary or secondary energy source is invoked when the \Nifs\, mass required is too high or the light curve is broad and long lasting. Both our light curve code and semi-analytic models for \Nifs\, decay as a primary energy source reproduce the early phase quite well without requiring significant amounts of synthesised \Nifs. Given that only the late time ( $t > 140$ d) shows deviation from a typical \Nifs\ decay model, a magnetar model would have to only show evidence in this late phase, or have a turnoff point close to day 140 which is difficult to explain.

\subsubsection{Time Dependent Gamma-ray Opacity}\label{subsubsec:gamma_op}
Allowing for a time dependant $\gamma$--ray opacity can reduce the total flux of the later two epochs that matches the multiplicative factor. This would also alter the bolometric light curve as less flux is trapped. However, we know of no mechanism that can arbitrarily change the $\gamma$--ray opacity over the time span of a month.


\section{Conclusions}\label{sec:conclu}

In this work, we model the light curve and the photospheric and nebular phase spectra to find a self consistent model for SN~2017ein. This model has an \mej\ of 1.6 $\pm$ 0.4 \msun, an \ek\ of 0.9 $\pm$ 0.2 foe, and a \mni\ of 0.09 $\pm$ 0.02 \msun.These parameters are required to reproduce the narrow lined spectra and compare favourably to a progenitor mass of 16--20 \msun.  The drop in the late time photometry and nebular flux is unexplained using our models, but the lack of spectroscopic data at \lam\ $>$ 8000 \AA, combined with the fading brightness of the object at late times makes the mechanism responsible for these deviations hard to determine. Common mechanisms that are often invoked to explain peculiar light curves do not seem to apply in this case, barring very exotic situations. 

Deriving a progenitor mass from the pre--supernova imaging results in an estimated mass in the 40-80 \msun\ range, depending on a single or binary star evolution or a compact cluster. Under reasonable assumptions for mass loss and evolution, this \mzams\ results in an \mej\ of 4--8 \msun. This is a strong contradiction between the modelled results and the observed estimations.

The evidence from the SN itself favours the model properties found in this work. As the progenitor and ejecta mass increase, the required binding energy that results in a successful explosion typically increases. The narrow lines and measured line velocities do not favour high energy models. High mass events also typically have broader light curves, which SN~2017ein also does not show. Similarly, as the \ek\ and \mej\ increases, the \Nifs\ mass synthesised typically increases. The \Nifs\ mass in this work does not match that of high energy and mass events. These results are all contradictory if one assumes a massive progenitor and a resulting large \mej. 

The pre--supernova observations show a slight offset in the supernova position and that of the point source. This offset is not significant, but combined with the contradictory results regarding explosion properties suggests the likely chance that the progenitor is not associated with the point like source and is a faint object within the line of sight. Exotic situations, such as a massive remnant with a low mass ejecta could link the two results, but is a fine tuned mechanism with weak explanation. Further Hubble imaging after the SN~2017ein fades will hopefully solve this problem.

\section*{Acknowledgements}
JT is funded by an STFC grant.
CA is supported by NASA grant 80NSSC19K1717 and NSF grants AST-1920392 and AST-1911074.
SJP is supported by H2020 ERC grant no.~758638.
The Liverpool Telescope is operated on the island of La Palma by Liverpool John Moores University in the Spanish Observatorio del Roque de los Muchachos of the Instituto de Astrofisica de Canarias with financial support from the UK Science and Technology Facilities Council.

\section*{Data Availability}
Data will be made available on the Weizmann Interactive Supernova Data Repository (WISeREP) at https://wiserep.weizmann.ac.il/.




\bibliographystyle{mnras}
\bibliography{bib} 

\begin{thebibliography}{}
\makeatletter
\relax
\def\mn@urlcharsother{\let\do\@makeother \do\$\do\&\do\#\do\^\do\_\do\%\do\~}
\def\mn@doi{\begingroup\mn@urlcharsother \@ifnextchar [ {\mn@doi@}
  {\mn@doi@[]}}
\def\mn@doi@[#1]#2{\def\@tempa{#1}\ifx\@tempa\@empty \href
  {http://dx.doi.org/#2} {doi:#2}\else \href {http://dx.doi.org/#2} {#1}\fi
  \endgroup}
\def\mn@eprint#1#2{\mn@eprint@#1:#2::\@nil}
\def\mn@eprint@arXiv#1{\href {http://arxiv.org/abs/#1} {{\tt arXiv:#1}}}
\def\mn@eprint@dblp#1{\href {http://dblp.uni-trier.de/rec/bibtex/#1.xml}
  {dblp:#1}}
\def\mn@eprint@#1:#2:#3:#4\@nil{\def\@tempa {#1}\def\@tempb {#2}\def\@tempc
  {#3}\ifx \@tempc \@empty \let \@tempc \@tempb \let \@tempb \@tempa \fi \ifx
  \@tempb \@empty \def\@tempb {arXiv}\fi \@ifundefined
  {mn@eprint@\@tempb}{\@tempb:\@tempc}{\expandafter \expandafter \csname
  mn@eprint@\@tempb\endcsname \expandafter{\@tempc}}}

\bibitem[\protect\citeauthoryear{{Ahn} et~al.,}{{Ahn} et~al.}{2012}]{Ahn2012}
{Ahn} C.~P.,  et~al., 2012, \mn@doi [\apjs] {10.1088/0067-0049/203/2/21}, \href
  {https://ui.adsabs.harvard.edu/abs/2012ApJS..203...21A} {203, 21}

\bibitem[\protect\citeauthoryear{{Arbour}}{{Arbour}}{2017}]{ArbourTNS}
{Arbour} R.,  2017, Transient Name Server Discovery Report, \href
  {https://ui.adsabs.harvard.edu/abs/2017TNSTR.588....1A} {2017-588, 1}

\bibitem[\protect\citeauthoryear{{Arnett}}{{Arnett}}{1982}]{Arnett1982}
{Arnett} W.~D.,  1982, \mn@doi [\apj] {10.1086/159681}, \href
  {http://adsabs.harvard.edu/abs/1982ApJ...253..785A} {253, 785}

\bibitem[\protect\citeauthoryear{Ashall \& Mazzali}{Ashall \&
  Mazzali}{2020}]{Ashall2020}
Ashall C.,  Mazzali P.~A.,  2020, \mn@doi [\mnras] {10.1093/mnras/staa212},
  492, 5956

\bibitem[\protect\citeauthoryear{{Ashall} et~al.,}{{Ashall}
  et~al.}{2019}]{Ashall2019}
{Ashall} C.,  et~al., 2019, \mn@doi [\mnras] {10.1093/mnras/stz1588}, \href
  {https://ui.adsabs.harvard.edu/abs/2019MNRAS.487.5824A} {487, 5824}

\bibitem[\protect\citeauthoryear{{Axelrod}}{{Axelrod}}{1980}]{Axelrod1980}
{Axelrod} T.~S.,  1980, in {Wheeler} J.~C.,  ed., Texas Workshop on Type I
  Supernovae. pp 80--95

\bibitem[\protect\citeauthoryear{{Barnsley}, {Smith}  \& {Steele}}{{Barnsley}
  et~al.}{2012}]{Barnsley2012}
{Barnsley} R.~M.,  {Smith} R.~J.,   {Steele} I.~A.,  2012, \mn@doi
  [Astronomische Nachrichten] {10.1002/asna.201111634}, \href
  {http://adsabs.harvard.edu/abs/2012AN....333..101B} {333, 101}

\bibitem[\protect\citeauthoryear{{Cappellaro}, {Mazzali}, {Benetti},
  {Danziger}, {Turatto}, {della Valle}  \& {Patat}}{{Cappellaro}
  et~al.}{1997}]{Cappellaro1997}
{Cappellaro} E.,  {Mazzali} P.~A.,  {Benetti} S.,  {Danziger} I.~J.,  {Turatto}
  M.,  {della Valle} M.,   {Patat} F.,  1997, \aap, \href
  {https://ui.adsabs.harvard.edu/abs/1997A&A...328..203C} {328, 203}

\bibitem[\protect\citeauthoryear{{Cardelli}, {Clayton}  \& {Mathis}}{{Cardelli}
  et~al.}{1989}]{CCM}
{Cardelli} J.~A.,  {Clayton} G.~C.,   {Mathis} J.~S.,  1989, \mn@doi [\apj]
  {10.1086/167900}, \href {http://adsabs.harvard.edu/abs/1989ApJ...345..245C}
  {345, 245}

\bibitem[\protect\citeauthoryear{{Choi}, {Dotter}, {Conroy}, {Cantiello},
  {Paxton}  \& {Johnson}}{{Choi} et~al.}{2016}]{mist}
{Choi} J.,  {Dotter} A.,  {Conroy} C.,  {Cantiello} M.,  {Paxton} B.,
  {Johnson} B.~D.,  2016, \mn@doi [\apj] {10.3847/0004-637X/823/2/102}, \href
  {https://ui.adsabs.harvard.edu/abs/2016ApJ...823..102C} {823, 102}

\bibitem[\protect\citeauthoryear{{Davies} \& {Beasor}}{{Davies} \&
  {Beasor}}{2018}]{Davies2018}
{Davies} B.,  {Beasor} E.~R.,  2018, \mn@doi [\mnras] {10.1093/mnras/stx2734},
  \href {https://ui.adsabs.harvard.edu/abs/2018MNRAS.474.2116D} {474, 2116}

\bibitem[\protect\citeauthoryear{{Davies} \& {Beasor}}{{Davies} \&
  {Beasor}}{2020}]{Davies2020}
{Davies} B.,  {Beasor} E.~R.,  2020, \mn@doi [\mnras] {10.1093/mnras/staa174},
  \href {https://ui.adsabs.harvard.edu/abs/2020MNRAS.tmp..170D} {}

\bibitem[\protect\citeauthoryear{{Dexter} \& {Kasen}}{{Dexter} \&
  {Kasen}}{2013}]{Dexter2013}
{Dexter} J.,  {Kasen} D.,  2013, \mn@doi [\apj] {10.1088/0004-637X/772/1/30},
  \href {https://ui.adsabs.harvard.edu/abs/2013ApJ...772...30D} {772, 30}

\bibitem[\protect\citeauthoryear{{Drout} et~al.,}{{Drout}
  et~al.}{2011}]{Drout2011}
{Drout} M.~R.,  et~al., 2011, \mn@doi [\apj] {10.1088/0004-637X/741/2/97},
  \href {http://adsabs.harvard.edu/abs/2011ApJ...741...97D} {741, 97}

\bibitem[\protect\citeauthoryear{{Eldridge}, {Stanway}, {Xiao}, {McClelland },
  {Taylor}, {Ng}, {Greis}  \& {Bray}}{{Eldridge} et~al.}{2017}]{bpass1}
{Eldridge} J.~J.,  {Stanway} E.~R.,  {Xiao} L.,  {McClelland } L.~A.~S.,
  {Taylor} G.,  {Ng} M.,  {Greis} S.~M.~L.,   {Bray} J.~C.,  2017, \mn@doi
  [\pasa] {10.1017/pasa.2017.51}, \href
  {https://ui.adsabs.harvard.edu/abs/2017PASA...34...58E} {34, e058}

\bibitem[\protect\citeauthoryear{{Elmhamdi}, {Danziger}, {Branch},
  {Leibundgut}, {Baron}  \& {Kirshner}}{{Elmhamdi} et~al.}{2006}]{Elmhamdi2006}
{Elmhamdi} A.,  {Danziger} I.~J.,  {Branch} D.,  {Leibundgut} B.,  {Baron} E.,
   {Kirshner} R.~P.,  2006, \mn@doi [\aap] {10.1051/0004-6361:20054366}, \href
  {http://adsabs.harvard.edu/abs/2006A%26A...450..305E} {450, 305}

\bibitem[\protect\citeauthoryear{Ertl, Woosley, Sukhbold  \& Janka}{Ertl
  et~al.}{2020}]{Ertl2020}
Ertl T.,  Woosley S.~E.,  Sukhbold T.,   Janka H.-T.,  2020, \mn@doi [\apj]
  {10.3847/1538-4357/ab6458}, 890, 51

\bibitem[\protect\citeauthoryear{{Filippenko} et~al.,}{{Filippenko}
  et~al.}{1995}]{Filippenko1995}
{Filippenko} A.~V.,  et~al., 1995, \mn@doi [\apjl] {10.1086/309659}, \href
  {http://adsabs.harvard.edu/abs/1995ApJ...450L..11F} {450, L11}

\bibitem[\protect\citeauthoryear{{Foley} et~al.,}{{Foley}
  et~al.}{2003}]{Foley2003}
{Foley} R.~J.,  et~al., 2003, \mn@doi [\pasp] {10.1086/378242}, \href
  {http://adsabs.harvard.edu/abs/2003PASP..115.1220F} {115, 1220}

\bibitem[\protect\citeauthoryear{Fryer, Belczynski, Wiktorowicz, Dominik,
  Kalogera  \& Holz}{Fryer et~al.}{2012}]{Fryer2012}
Fryer C.~L.,  Belczynski K.,  Wiktorowicz G.,  Dominik M.,  Kalogera V.,   Holz
  D.~E.,  2012, \mn@doi [\apj] {10.1088/0004-637x/749/1/91}, 749, 91

\bibitem[\protect\citeauthoryear{{Georgy}, {Ekstr{\"o}m}, {Meynet}, {Massey},
  {Levesque}, {Hirschi}, {Eggenberger}  \& {Maeder}}{{Georgy}
  et~al.}{2012}]{georgy2012}
{Georgy} C.,  {Ekstr{\"o}m} S.,  {Meynet} G.,  {Massey} P.,  {Levesque} E.~M.,
  {Hirschi} R.,  {Eggenberger} P.,   {Maeder} A.,  2012, \mn@doi [\aap]
  {10.1051/0004-6361/201118340}, \href
  {http://adsabs.harvard.edu/abs/2012A%26A...542A..29G} {542, A29}

\bibitem[\protect\citeauthoryear{{Gottlieb}, {Nakar}  \& {Bromberg}}{{Gottlieb}
  et~al.}{2021}]{Gottlieb2021}
{Gottlieb} O.,  {Nakar} E.,   {Bromberg} O.,  2021, \mn@doi [\mnras]
  {10.1093/mnras/staa3501}, \href
  {https://ui.adsabs.harvard.edu/abs/2021MNRAS.500.3511G} {500, 3511}

\bibitem[\protect\citeauthoryear{{Hachinger}, {Mazzali}, {Taubenberger},
  {Pakmor}  \& {Hillebrandt}}{{Hachinger} et~al.}{2009}]{Hachinger2009}
{Hachinger} S.,  {Mazzali} P.~A.,  {Taubenberger} S.,  {Pakmor} R.,
  {Hillebrandt} W.,  2009, \mn@doi [\mnras] {10.1111/j.1365-2966.2009.15403.x},
  \href {http://adsabs.harvard.edu/abs/2009MNRAS.399.1238H} {399, 1238}

\bibitem[\protect\citeauthoryear{{Hachinger}, {Mazzali}, {Taubenberger},
  {Hillebrandt}, {Nomoto}  \& {Sauer}}{{Hachinger}
  et~al.}{2012}]{Hachinger2012}
{Hachinger} S.,  {Mazzali} P.~A.,  {Taubenberger} S.,  {Hillebrandt} W.,
  {Nomoto} K.,   {Sauer} D.~N.,  2012, \mn@doi [\mnras]
  {10.1111/j.1365-2966.2012.20464.x}, \href
  {http://adsabs.harvard.edu/abs/2012MNRAS.422...70H} {422, 70}

\bibitem[\protect\citeauthoryear{{Hunter} et~al.,}{{Hunter}
  et~al.}{2009}]{Hunter2009}
{Hunter} D.~J.,  et~al., 2009, \mn@doi [\aap] {10.1051/0004-6361/200912896},
  \href {https://ui.adsabs.harvard.edu/abs/2009A&A...508..371H} {508, 371}

\bibitem[\protect\citeauthoryear{{Kilpatrick} et~al.,}{{Kilpatrick}
  et~al.}{2018}]{Kilpatrick2018}
{Kilpatrick} C.~D.,  et~al., 2018, \mn@doi [\mnras] {10.1093/mnras/sty2022},
  \href {https://ui.adsabs.harvard.edu/abs/2018MNRAS.480.2072K} {480, 2072}

\bibitem[\protect\citeauthoryear{{Kozma} \& {Fransson}}{{Kozma} \&
  {Fransson}}{1998}]{Kozma1998}
{Kozma} C.,  {Fransson} C.,  1998, \mn@doi [\apj] {10.1086/305409}, \href
  {https://ui.adsabs.harvard.edu/abs/1998ApJ...496..946K} {496, 946}

\bibitem[\protect\citeauthoryear{{Langer}}{{Langer}}{2012}]{Langer2012}
{Langer} N.,  2012, \mn@doi [\araa] {10.1146/annurev-astro-081811-125534},
  \href {http://adsabs.harvard.edu/abs/2012ARA%26A..50..107L} {50, 107}

\bibitem[\protect\citeauthoryear{{Lucy}}{{Lucy}}{1999}]{Lucy1999}
{Lucy} L.~B.,  1999, \aap, \href
  {http://adsabs.harvard.edu/abs/1999A%26A...345..211L} {345, 211}

\bibitem[\protect\citeauthoryear{{Lyman}, {Bersier}, {James}, {Mazzali},
  {Eldridge}, {Fraser}  \& {Pian}}{{Lyman} et~al.}{2016}]{Lyman2016}
{Lyman} J.~D.,  {Bersier} D.,  {James} P.~A.,  {Mazzali} P.~A.,  {Eldridge}
  J.~J.,  {Fraser} M.,   {Pian} E.,  2016, \mn@doi [\mnras]
  {10.1093/mnras/stv2983}, \href
  {https://ui.adsabs.harvard.edu/abs/2016MNRAS.457..328L} {457, 328}

\bibitem[\protect\citeauthoryear{{Maeda} et~al.,}{{Maeda}
  et~al.}{2008}]{Maeda2008}
{Maeda} K.,  et~al., 2008, \mn@doi [Science] {10.1126/science.1149437}, \href
  {http://adsabs.harvard.edu/abs/2008Sci...319.1220M} {319, 1220}

\bibitem[\protect\citeauthoryear{{Maund} \& {Ramirez-Ruiz}}{{Maund} \&
  {Ramirez-Ruiz}}{2016}]{Maund2016}
{Maund} J.~R.,  {Ramirez-Ruiz} E.,  2016, \mn@doi [\mnras]
  {10.1093/mnras/stv2760}, \href
  {https://ui.adsabs.harvard.edu/abs/2016MNRAS.456.3175M} {456, 3175}

\bibitem[\protect\citeauthoryear{{Mazzali}}{{Mazzali}}{2000}]{Mazzali2000b}
{Mazzali} P.~A.,  2000, \aap, \href
  {http://adsabs.harvard.edu/abs/2000A%26A...363..705M} {363, 705}

\bibitem[\protect\citeauthoryear{{Mazzali} \& {Lucy}}{{Mazzali} \&
  {Lucy}}{1993}]{Mazzali1993}
{Mazzali} P.~A.,  {Lucy} L.~B.,  1993, \aap, \href
  {http://adsabs.harvard.edu/abs/1993A%26A...279..447M} {279, 447}

\bibitem[\protect\citeauthoryear{{Mazzali}, {Lucy}  \& {Butler}}{{Mazzali}
  et~al.}{1992}]{Mazzali1992}
{Mazzali} P.~A.,  {Lucy} L.~B.,   {Butler} K.,  1992, \aap, \href
  {https://ui.adsabs.harvard.edu/abs/1992A&A...258..399M} {258, 399}

\bibitem[\protect\citeauthoryear{{Mazzali}, {Nomoto}, {Patat}  \&
  {Maeda}}{{Mazzali} et~al.}{2001}]{Mazzali2001}
{Mazzali} P.~A.,  {Nomoto} K.,  {Patat} F.,   {Maeda} K.,  2001, \mn@doi [\apj]
  {10.1086/322420}, \href {http://adsabs.harvard.edu/abs/2001ApJ...559.1047M}
  {559, 1047}

\bibitem[\protect\citeauthoryear{{Mazzali} et~al.,}{{Mazzali}
  et~al.}{2005}]{Mazzali2005}
{Mazzali} P.~A.,  et~al., 2005, \mn@doi [Science] {10.1126/science.1111384},
  \href {http://adsabs.harvard.edu/abs/2005Sci...308.1284M} {308, 1284}

\bibitem[\protect\citeauthoryear{{Mazzali} et~al.,}{{Mazzali}
  et~al.}{2007}]{Mazzali2007}
{Mazzali} P.~A.,  et~al., 2007, \mn@doi [\apj] {10.1086/521873}, \href
  {http://adsabs.harvard.edu/abs/2007ApJ...670..592M} {670, 592}

\bibitem[\protect\citeauthoryear{{Mazzali} et~al.,}{{Mazzali}
  et~al.}{2008}]{Mazzali2008}
{Mazzali} P.~A.,  et~al., 2008, \mn@doi [Science] {10.1126/science.1158088},
  \href {http://adsabs.harvard.edu/abs/2008Sci...321.1185M} {321, 1185}

\bibitem[\protect\citeauthoryear{{Mazzali}, {Deng}, {Hamuy}  \&
  {Nomoto}}{{Mazzali} et~al.}{2009}]{Mazzali2009}
{Mazzali} P.~A.,  {Deng} J.,  {Hamuy} M.,   {Nomoto} K.,  2009, \mn@doi [\apj]
  {10.1088/0004-637X/703/2/1624}, \href
  {http://adsabs.harvard.edu/abs/2009ApJ...703.1624M} {703, 1624}

\bibitem[\protect\citeauthoryear{{Mazzali}, {Maurer}, {Valenti}, {Kotak}  \&
  {Hunter}}{{Mazzali} et~al.}{2010}]{Mazzali2010}
{Mazzali} P.~A.,  {Maurer} I.,  {Valenti} S.,  {Kotak} R.,   {Hunter} D.,
  2010, \mn@doi [\mnras] {10.1111/j.1365-2966.2010.17133.x}, \href
  {http://adsabs.harvard.edu/abs/2010MNRAS.408...87M} {408, 87}

\bibitem[\protect\citeauthoryear{{Mazzali}, {Sauer}, {Pian}, {Deng},
  {Prentice}, {Ben Ami}, {Taubenberger}  \& {Nomoto}}{{Mazzali}
  et~al.}{2017}]{Mazzali2017}
{Mazzali} P.~A.,  {Sauer} D.~N.,  {Pian} E.,  {Deng} J.,  {Prentice} S.,  {Ben
  Ami} S.,  {Taubenberger} S.,   {Nomoto} K.,  2017, \mn@doi [\mnras]
  {10.1093/mnras/stx992}, \href
  {http://adsabs.harvard.edu/abs/2017MNRAS.469.2498M} {469, 2498}

\bibitem[\protect\citeauthoryear{{Mazzali}, {Ashall}, {Pian}, {Stritzinger},
  {Gall}, {Phillips}, {H{\"o}flich}  \& {Hsiao}}{{Mazzali}
  et~al.}{2018}]{Mazzali2018}
{Mazzali} P.~A.,  {Ashall} C.,  {Pian} E.,  {Stritzinger} M.~D.,  {Gall} C.,
  {Phillips} M.~M.,  {H{\"o}flich} P.,   {Hsiao} E.,  2018, \mn@doi [\mnras]
  {10.1093/mnras/sty434}, \href
  {https://ui.adsabs.harvard.edu/abs/2018MNRAS.476.2905M} {476, 2905}

\bibitem[\protect\citeauthoryear{{Mazzali}, {Moriya}, {Tanaka}  \&
  {Woosley}}{{Mazzali} et~al.}{2019}]{Mazzali2019}
{Mazzali} P.~A.,  {Moriya} T.~J.,  {Tanaka} M.,   {Woosley} S.~E.,  2019,
  \mn@doi [\mnras] {10.1093/mnras/stz177}, \href
  {https://ui.adsabs.harvard.edu/abs/2019MNRAS.484.3451M} {484, 3451}

\bibitem[\protect\citeauthoryear{{Mazzali} et~al.,}{{Mazzali}
  et~al.}{2020}]{Mazzali2020}
{Mazzali} P.~A.,  et~al., 2020, \mn@doi [\mnras] {10.1093/mnras/staa839}, \href
  {https://ui.adsabs.harvard.edu/abs/2020MNRAS.494.2809M} {494, 2809}

\bibitem[\protect\citeauthoryear{{Millard} et~al.,}{{Millard}
  et~al.}{1999}]{Millard1999}
{Millard} J.,  et~al., 1999, \mn@doi [\apj] {10.1086/308108}, \href
  {https://ui.adsabs.harvard.edu/abs/1999ApJ...527..746M} {527, 746}

\bibitem[\protect\citeauthoryear{{Nakar} \& {Piran}}{{Nakar} \&
  {Piran}}{2017}]{Nakar2017}
{Nakar} E.,  {Piran} T.,  2017, \mn@doi [\apj] {10.3847/1538-4357/834/1/28},
  \href {https://ui.adsabs.harvard.edu/abs/2017ApJ...834...28N} {834, 28}

\bibitem[\protect\citeauthoryear{{Nomoto}, {Yamaoka}, {Pols}, {van den Heuvel},
  {Iwamoto}, {Kumagai}  \& {Shigeyama}}{{Nomoto} et~al.}{1994}]{Nomoto1994}
{Nomoto} K.,  {Yamaoka} H.,  {Pols} O.~R.,  {van den Heuvel} E.~P.~J.,
  {Iwamoto} K.,  {Kumagai} S.,   {Shigeyama} T.,  1994, \mn@doi [\nat]
  {10.1038/371227a0}, \href {http://adsabs.harvard.edu/abs/1994Natur.371..227N}
  {371, 227}

\bibitem[\protect\citeauthoryear{Nomoto, Iwamoto  \& Suzuki}{Nomoto
  et~al.}{1995}]{Nomoto1995}
Nomoto K.,  Iwamoto K.,   Suzuki T.,  1995, \mn@doi [Physics Reports]
  {https://doi.org/10.1016/0370-1573(94)00107-E}, 256, 173

\bibitem[\protect\citeauthoryear{{Piascik}, {Steele}, {Bates}, {Mottram},
  {Smith}, {Barnsley}  \& {Bolton}}{{Piascik} et~al.}{2014}]{Piascik2014}
{Piascik} A.~S.,  {Steele} I.~A.,  {Bates} S.~D.,  {Mottram} C.~J.,  {Smith}
  R.~J.,  {Barnsley} R.~M.,   {Bolton} B.,  2014, in Ground-based and Airborne
  Instrumentation for Astronomy V. p. 91478H, \mn@doi{10.1117/12.2055117}

\bibitem[\protect\citeauthoryear{{Piran}, {Nakar}, {Mazzali}  \&
  {Pian}}{{Piran} et~al.}{2019}]{Piran2019}
{Piran} T.,  {Nakar} E.,  {Mazzali} P.,   {Pian} E.,  2019, \mn@doi [\apjl]
  {10.3847/2041-8213/aaffce}, \href
  {https://ui.adsabs.harvard.edu/abs/2019ApJ...871L..25P} {871, L25}

\bibitem[\protect\citeauthoryear{{Pols} \& {Dewi}}{{Pols} \&
  {Dewi}}{2002}]{Pols2002}
{Pols} O.~R.,  {Dewi} J. D.~M.,  2002, \mn@doi [\pasa] {10.1071/AS01121}, \href
  {https://ui.adsabs.harvard.edu/abs/2002PASA...19..233P} {19, 233}

\bibitem[\protect\citeauthoryear{{Poznanski}, {Prochaska}  \&
  {Bloom}}{{Poznanski} et~al.}{2012}]{Poznanski2012}
{Poznanski} D.,  {Prochaska} J.~X.,   {Bloom} J.~S.,  2012, \mn@doi [\mnras]
  {10.1111/j.1365-2966.2012.21796.x}, \href
  {http://adsabs.harvard.edu/abs/2012MNRAS.426.1465P} {426, 1465}

\bibitem[\protect\citeauthoryear{{Prentice} \& {Mazzali}}{{Prentice} \&
  {Mazzali}}{2017}]{Prentice2017}
{Prentice} S.~J.,  {Mazzali} P.~A.,  2017, \mn@doi [\mnras]
  {10.1093/mnras/stx980}, \href
  {http://adsabs.harvard.edu/abs/2017MNRAS.469.2672P} {469, 2672}

\bibitem[\protect\citeauthoryear{{Prentice} et~al.,}{{Prentice}
  et~al.}{2016}]{Prentice2016}
{Prentice} S.~J.,  et~al., 2016, \mn@doi [\mnras] {10.1093/mnras/stw299}, \href
  {http://adsabs.harvard.edu/abs/2016MNRAS.458.2973P} {458, 2973}

\bibitem[\protect\citeauthoryear{{Prentice} et~al.,}{{Prentice}
  et~al.}{2018}]{Prentice2018}
{Prentice} S.~J.,  et~al., 2018, \mn@doi [\mnras] {10.1093/mnras/sty1223},
  \href {https://ui.adsabs.harvard.edu/abs/2018MNRAS.478.4162P} {478, 4162}

\bibitem[\protect\citeauthoryear{{Prentice} et~al.,}{{Prentice}
  et~al.}{2019}]{Prentice2019}
{Prentice} S.~J.,  et~al., 2019, \mn@doi [\mnras] {10.1093/mnras/sty3399},
  \href {https://ui.adsabs.harvard.edu/abs/2019MNRAS.485.1559P} {485, 1559}

\bibitem[\protect\citeauthoryear{{Sauer}, {Mazzali}, {Deng}, {Valenti},
  {Nomoto}  \& {Filippenko}}{{Sauer} et~al.}{2006}]{Sauer2006}
{Sauer} D.~N.,  {Mazzali} P.~A.,  {Deng} J.,  {Valenti} S.,  {Nomoto} K.,
  {Filippenko} A.~V.,  2006, \mn@doi [\mnras]
  {10.1111/j.1365-2966.2006.10438.x}, \href
  {http://adsabs.harvard.edu/abs/2006MNRAS.369.1939S} {369, 1939}

\bibitem[\protect\citeauthoryear{{Schlafly} \& {Finkbeiner}}{{Schlafly} \&
  {Finkbeiner}}{2011}]{Schlafly2011}
{Schlafly} E.~F.,  {Finkbeiner} D.~P.,  2011, \mn@doi [\apj]
  {10.1088/0004-637X/737/2/103}, \href
  {http://adsabs.harvard.edu/abs/2011ApJ...737..103S} {737, 103}

\bibitem[\protect\citeauthoryear{{Smartt}}{{Smartt}}{2009}]{Smartt2009}
{Smartt} S.~J.,  2009, \mn@doi [\araa] {10.1146/annurev-astro-082708-101737},
  \href {http://adsabs.harvard.edu/abs/2009ARA%26A..47...63S} {47, 63}

\bibitem[\protect\citeauthoryear{{Smartt}, {Eldridge}, {Crockett}  \&
  {Maund}}{{Smartt} et~al.}{2009}]{Smartt20092}
{Smartt} S.~J.,  {Eldridge} J.~J.,  {Crockett} R.~M.,   {Maund} J.~R.,  2009,
  \mn@doi [\mnras] {10.1111/j.1365-2966.2009.14506.x}, \href
  {https://ui.adsabs.harvard.edu/abs/2009MNRAS.395.1409S} {395, 1409}

\bibitem[\protect\citeauthoryear{{Steele} et~al.,}{{Steele}
  et~al.}{2004}]{Steele2004}
{Steele} I.~A.,  et~al., 2004, in {Oschmann} Jr. J.~M.,  ed.,  \procspie Vol.
  5489, Ground-based Telescopes. pp 679--692, \mn@doi{10.1117/12.551456}

\bibitem[\protect\citeauthoryear{{Stehle}, {Mazzali}, {Benetti}  \&
  {Hillebrandt}}{{Stehle} et~al.}{2005}]{Stehle2005}
{Stehle} M.,  {Mazzali} P.~A.,  {Benetti} S.,   {Hillebrandt} W.,  2005,
  \mn@doi [\mnras] {10.1111/j.1365-2966.2005.09116.x}, \href
  {http://ukads.nottingham.ac.uk/abs/2005MNRAS.360.1231S} {360, 1231}

\bibitem[\protect\citeauthoryear{{Taddia} et~al.,}{{Taddia}
  et~al.}{2018}]{Taddia2018}
{Taddia} F.,  et~al., 2018, \mn@doi [\aap] {10.1051/0004-6361/201730844}, \href
  {https://ui.adsabs.harvard.edu/abs/2018A&A...609A.136T} {609, A136}

\bibitem[\protect\citeauthoryear{{Taubenberger} et~al.,}{{Taubenberger}
  et~al.}{2006}]{Taubenberger2006}
{Taubenberger} S.,  et~al., 2006, \mn@doi [\mnras]
  {10.1111/j.1365-2966.2006.10776.x}, \href
  {http://adsabs.harvard.edu/abs/2006MNRAS.371.1459T} {371, 1459}

\bibitem[\protect\citeauthoryear{{Taubenberger} et~al.,}{{Taubenberger}
  et~al.}{2009}]{Taubenberger2009}
{Taubenberger} S.,  et~al., 2009, \mn@doi [\mnras]
  {10.1111/j.1365-2966.2009.15003.x}, \href
  {http://adsabs.harvard.edu/abs/2009MNRAS.397..677T} {397, 677}

\bibitem[\protect\citeauthoryear{{Teffs}, {Ertl}, {Mazzali}, {Hachinger}  \&
  {Janka}}{{Teffs} et~al.}{2020a}]{Teffs2020}
{Teffs} J.,  {Ertl} T.,  {Mazzali} P.,  {Hachinger} S.,   {Janka} T.,  2020a,
  \mn@doi [\mnras] {10.1093/mnras/staa123}, \href
  {https://ui.adsabs.harvard.edu/abs/2020MNRAS.tmp..116T} {p.~116}

\bibitem[\protect\citeauthoryear{{Teffs}, {Ertl}, {Mazzali}, {Hachinger}  \&
  {Janka}}{{Teffs} et~al.}{2020b}]{Teffs2020b}
{Teffs} J.,  {Ertl} T.,  {Mazzali} P.,  {Hachinger} S.,   {Janka} H.~T.,
  2020b, arXiv e-prints, \href
  {https://ui.adsabs.harvard.edu/abs/2020arXiv200807887T} {p. arXiv:2008.07887}

\bibitem[\protect\citeauthoryear{{Tomasella} et~al.,}{{Tomasella}
  et~al.}{2013}]{Tomasella2013}
{Tomasella} L.,  et~al., 2013, \mn@doi [\mnras] {10.1093/mnras/stt1130}, \href
  {https://ui.adsabs.harvard.edu/abs/2013MNRAS.434.1636T} {434, 1636}

\bibitem[\protect\citeauthoryear{{Valenti} et~al.,}{{Valenti}
  et~al.}{2008}]{Valenti2008_07gr}
{Valenti} S.,  et~al., 2008, \mn@doi [\apjl] {10.1086/527672}, \href
  {https://ui.adsabs.harvard.edu/abs/2008ApJ...673L.155V} {673, L155}

\bibitem[\protect\citeauthoryear{{Valenti} et~al.,}{{Valenti}
  et~al.}{2012}]{Valenti2012}
{Valenti} S.,  et~al., 2012, \mn@doi [\apjl] {10.1088/2041-8205/749/2/L28},
  \href {http://adsabs.harvard.edu/abs/2012ApJ...749L..28V} {749, L28}

\bibitem[\protect\citeauthoryear{{Van Dyk} et~al.,}{{Van Dyk}
  et~al.}{2014}]{VanDyk2014}
{Van Dyk} S.~D.,  et~al., 2014, \mn@doi [\aj] {10.1088/0004-6256/147/2/37},
  \href {http://adsabs.harvard.edu/abs/2014AJ....147...37V} {147, 37}

\bibitem[\protect\citeauthoryear{{Van Dyk} et~al.,}{{Van Dyk}
  et~al.}{2018}]{VanDyk2018}
{Van Dyk} S.~D.,  et~al., 2018, \mn@doi [\apj] {10.3847/1538-4357/aac32c},
  \href {https://ui.adsabs.harvard.edu/abs/2018ApJ...860...90V} {860, 90}

\bibitem[\protect\citeauthoryear{{Wiktorowicz}, {Belczynski}  \&
  {Maccarone}}{{Wiktorowicz} et~al.}{2014}]{Wiktorowicz2014}
{Wiktorowicz} G.,  {Belczynski} K.,   {Maccarone} T.,  2014, in {de Grijs} R.,
  ed., Binary Systems, their Evolution and Environments. p.~37 (\mn@eprint
  {arXiv} {1312.5924})

\bibitem[\protect\citeauthoryear{Woosley}{Woosley}{2019}]{Woosley2019}
Woosley S.~E.,  2019, \mn@doi [\apj] {10.3847/1538-4357/ab1b41}, 878, 49

\bibitem[\protect\citeauthoryear{{Woosley} \& {Weaver}}{{Woosley} \&
  {Weaver}}{1995}]{Woosley1995}
{Woosley} S.~E.,  {Weaver} T.~A.,  1995, \mn@doi [\apjs] {10.1086/192237},
  \href {https://ui.adsabs.harvard.edu/abs/1995ApJS..101..181W} {101, 181}

\bibitem[\protect\citeauthoryear{{Woosley}, {Langer}  \& {Weaver}}{{Woosley}
  et~al.}{1993}]{Woosley_Langer1993}
{Woosley} S.~E.,  {Langer} N.,   {Weaver} T.~A.,  1993, \mn@doi [\apj]
  {10.1086/172886}, \href
  {https://ui.adsabs.harvard.edu/abs/1993ApJ...411..823W} {411, 823}

\bibitem[\protect\citeauthoryear{{Woosley}, {Langer}  \& {Weaver}}{{Woosley}
  et~al.}{1995}]{Woosley_Langer1995}
{Woosley} S.~E.,  {Langer} N.,   {Weaver} T.~A.,  1995, \mn@doi [\apj]
  {10.1086/175963}, \href
  {https://ui.adsabs.harvard.edu/abs/1995ApJ...448..315W} {448, 315}

\bibitem[\protect\citeauthoryear{{Woosley}, {Sukhbold}  \& {Kasen}}{{Woosley}
  et~al.}{2020}]{Woosley2020}
{Woosley} S.,  {Sukhbold} T.,   {Kasen} D.,  2020, arXiv e-prints, \href
  {https://ui.adsabs.harvard.edu/abs/2020arXiv200906868W} {p. arXiv:2009.06868}

\bibitem[\protect\citeauthoryear{{Xiang}, {Rui}, {Wang}, {Song}, {Xiao},
  {Zhang}  \& {Zhang}}{{Xiang} et~al.}{2017}]{2017TNSCR.599....1X}
{Xiang} D.,  {Rui} L.,  {Wang} X.,  {Song} H.,  {Xiao} F.,  {Zhang} T.,
  {Zhang} J.,  2017, Transient Name Server Classification Report, \href
  {https://ui.adsabs.harvard.edu/abs/2017TNSCR.599....1X} {2017-599, 1}

\bibitem[\protect\citeauthoryear{{Xiang} et~al.,}{{Xiang}
  et~al.}{2019}]{Xiang2019}
{Xiang} D.,  et~al., 2019, \mn@doi [\apj] {10.3847/1538-4357/aaf8b0}, \href
  {https://ui.adsabs.harvard.edu/abs/2019ApJ...871..176X} {871, 176}

\bibitem[\protect\citeauthoryear{Yoon}{Yoon}{2017}]{Yoon2017}
Yoon S.-C.,  2017, \mn@doi [\mnras] {10.1093/mnras/stx1496}, 470, 3970

\bibitem[\protect\citeauthoryear{Yoon, Woosley  \& Langer}{Yoon
  et~al.}{2010}]{Yoon2010}
Yoon S.-C.,  Woosley S.~E.,   Langer N.,  2010, \mn@doi [\apj]
  {10.1088/0004-637x/725/1/940}, 725, 940

\makeatother
\end{thebibliography}




\appendix


\bsp	
\label{lastpage}
\end{document}